\documentclass[nohyper,12pt,letterpaper]{JHEP3}
\usepackage{epsfig}
\usepackage[latin1]{inputenc}
\usepackage{bbm,amsfonts}
\usepackage{graphicx}
\usepackage{amssymb,amsmath}
\usepackage{fancybox}

\author{Marco S. Bianchi$^\ast$,
  Gaston Giribet$^\hash$,
  Matias Leoni$^{\hash}$,  
  and Silvia Penati$\dag$\\\\
  $^\ast$Institut f\"ur Physik,
Humboldt-Universit\"at zu Berlin,
Newtonstra{\ss}e 15, 12489 Berlin, Germany \\\\
  $^\hash$Physics Department, FCEyN-UBA \& IFIBA-CONICET\
Ciudad Universitaria, Pabell\'on I, 1428, Buenos Aires, Argentina \\\\
  $^\dag$Dipartimento di Fisica, Universit\`a degli studi di Milano--Bicocca and
  INFN, Sezione di Milano--Bicocca, Piazza della Scienza 3, I-20126 Milano, Italy \\
  \qquad\\
  E-mail: \email{ marco.bianchi@physik.hu-berlin.de, gaston@df.uba.ar, 
  leoni@df.uba.ar,
    silvia.penati@mib.infn.it}
}

\abstract{We compute the expectation value of the 1/2 BPS circular Wilson loop operator in ABJ(M) theory at two loops in perturbation theory.
Our result turns out to be in exact agreement with the weak coupling limit of the prediction coming from localization, including finite N contributions associated to non--planar diagrams. It also confirms the identification of the correct framing factor that connects framing-zero and framing-one expressions, previously proposed. 
The evaluation of the 1/2 BPS operator is made technically difficult in comparison with other observables of ABJ(M) theory by the appearance of integrals involving the coupling between fermions and gauge fields, which are absent for instance in the 1/6 BPS case. We describe in detail how to analytically solve these integrals in dimensional regularization with dimensional reduction (DRED). By suitably performing the physical limit to three dimensions we clarify the role played by short distance divergences on the final result and the mechanism of their cancellation. }

\preprint{July 2013\\HU-EP-13/30}

\title{The 1/2 BPS Wilson loop in ABJ(M) at two loops: The details}


\csname @addtoreset\endcsname{equation}{section}

\def\bseq{\begin{subequation}}  
\def\eseq{\end{subequation}}
\def\bsea{\begin{subeqnarray}}  
\def\esea{\end{subeqnarray}}

\newcommand{\beq}{\begin{equation}}
\newcommand{\bea}{\begin{eqnarray}}
\newcommand{\eea}{\end{eqnarray}}
\newcommand{\eeq}{\end{equation}}

\newcommand {\non}{\nonumber}

\renewcommand{\a}{\alpha}
\renewcommand{\b}{\beta}

\renewcommand{\d}{\delta}
\newcommand{\pa}{\partial}
\newcommand{\g}{\gamma}
\newcommand{\G}{\Gamma}

\newcommand{\e}{\epsilon}

\renewcommand{\l}{\lambda}

\newcommand{\s}{\sigma}

\renewcommand{\t}{\tau}

\def\Mb{\kern 2pt\mathchoice
        {
         \vbox{\hrule width10pt height 0.4pt depth 0pt
         \kern 1.2pt\hbox{\kern -2pt$\displaystyle M$}}}
        {
         \vbox{\hrule width10pt height 0.4pt depth 0pt
         \kern 1.2pt\hbox{\kern -2pt$\textstyle M$}}}
        {
\vbox{\hrule width6pt height 0.4pt depth 0pt
         \kern 1.0pt\hbox{\kern -2pt$\scriptstyle M$}}}
        {
         \vbox{\hrule width5pt height 0.4pt depth 0pt
         \kern 0.8pt\hbox{\kern -2pt$\scriptscriptstyle M$}}}}

\def\Sb{\kern 2pt\mathchoice
        {
         \vbox{\hrule width6pt height 0.4pt depth 0pt
         \kern 1.2pt\hbox{\kern -2pt$\displaystyle S$}}}
        {
         \vbox{\hrule width6pt height 0.4pt depth 0pt
         \kern 1.2pt\hbox{\kern -2pt$\textstyle S$}}}
        {
         \vbox{\hrule width3.5pt height 0.4pt depth 0pt
         \kern 1.0pt\hbox{\kern -2pt$\scriptstyle S$}}}
        {
         \vbox{\hrule width3pt height 0.4pt depth 0pt
         \kern 0.8pt\hbox{\kern -2pt$\scriptscriptstyle S$}}}}

\def\Rb{\kern 2pt\mathchoice
        {
         \vbox{\hrule width5.5pt height 0.4pt depth 0pt
         \kern 1.2pt\hbox{\kern -2.5pt$\displaystyle R$}}}
        {
         \vbox{\hrule width5.5pt height 0.4pt depth 0pt
         \kern 1.2pt\hbox{\kern -2.5pt$\textstyle R$}}}
        {
         \vbox{\hrule width3.5pt height 0.4pt depth 0pt
         \kern 1.0pt\hbox{\kern -2.2pt$\scriptstyle R$}}}
        {
         \vbox{\hrule width3pt height 0.4pt depth 0pt
         \kern 0.8pt\hbox{\kern -2.2pt$\scriptscriptstyle R$}}}}

  \def\pp{{\mathchoice
      {
          \kern 1pt%
          \raise 1pt
          \vbox{\hrule width5pt height0.4pt depth0pt
            \kern -2pt
            \hbox{\kern 2.3pt
              \vrule width0.4pt height6pt depth0pt
              }
            \kern -2pt
            \hrule width5pt height0.4pt depth0pt}%
            \kern 1pt
       }
        {
          \kern 1pt%
          \raise 1pt
          \vbox{\hrule width4.3pt height0.4pt depth0pt
            \kern -1.8pt
            \hbox{\kern 1.95pt
              \vrule width0.4pt height5.4pt depth0pt
              }
            \kern -1.8pt
            \hrule width4.3pt height0.4pt depth0pt}%
            \kern 1pt
        }
        {
          \kern 0.5pt%
          \raise 1pt
          \vbox{\hrule width4.0pt height0.3pt depth0pt
            \kern -1.9pt  
            \hbox{\kern 1.85pt
              \vrule width0.3pt height5.7pt depth0pt
              }
            \kern -1.9pt
            \hrule width4.0pt height0.3pt depth0pt}%
            \kern 0.5pt
        }
        {
          \kern 0.5pt%
          \raise 1pt
          \vbox{\hrule width3.6pt height0.3pt depth0pt
            \kern -1.5pt
            \hbox{\kern 1.65pt
              \vrule width0.3pt height4.5pt depth0pt
              }
            \kern -1.5pt
            \hrule width3.6pt height0.3pt depth0pt}%
            \kern 0.5pt
        }
    }}

  \def\mm{{\mathchoice
               {
                 \kern 1pt
           \raise 1pt    \vbox{\hrule width5pt height0.4pt depth0pt
                  \kern 2pt
                  \hrule width5pt height0.4pt depth0pt}
                 \kern 1pt}
               {
                \kern 1pt
           \raise 1pt \vbox{\hrule width4.3pt height0.4pt depth0pt
                  \kern 1.8pt
                  \hrule width4.3pt height0.4pt depth0pt}
                 \kern 1pt}
               {
                \kern 0.5pt
           \raise 1pt
                \vbox{\hrule width4.0pt height0.3pt depth0pt
                  \kern 1.9pt
                  \hrule width4.0pt height0.3pt depth0pt}
                \kern 1pt}
               {
               \kern 0.5pt
         \raise 1pt  \vbox{\hrule width3.6pt height0.3pt depth0pt
                  \kern 1.5pt
                  \hrule width3.6pt height0.3pt depth0pt}
               \kern 0.5pt}
               }}

\def\pd{{\kern0.5pt
           + \kern-5.05pt \raise5.8pt\hbox{$\textstyle.$}\kern
0.5pt}}

\def\pmd{{\kern0.5pt
          \pm \kern-5.05pt
\raise6.3pt\hbox{$\textstyle.$}\kern1.5pt}}

\def\md{{\mathchoice
   {
      {{\kern 1pt - \kern-6.2pt \raise5pt\hbox{$\textstyle.$}\kern
1pt}}}
    {
      {{\kern 1pt - \kern-6.2pt \raise5pt\hbox{$\textstyle.$}\kern
1pt}}}
    {
      {\kern0.5pt - \kern-5.05pt
\raise3.4pt\hbox{$\textstyle.$}\kern0.5pt}}
    {
      {\kern0.5pt - \kern-5.05pt
\raise3.4pt\hbox{$\textstyle.$}\kern0.5pt}}}}

\def\beq{\begin{equation}}
\def\eeq{\end{equation}}
\def\bea{\begin{eqnarray}}
\def\eea{\end{eqnarray}}
\def\Tr{\textstyle{Tr}}

\def\a{\alpha}
\def\b{\beta}
\def\g{\gamma}

\def\d{\delta}
\def\e{\epsilon}

\def\l{\lambda}
\def\G{\Gamma}

\begin{document}

\section{Introduction}

The study of Wilson loops is of central importance in gauge theories since they contain relevant information, for instance about the potential between colored particles and about the Schwinger pair production probability. 
For supersymmetric gauge theories Wilson loop operators can be defined, which preserve a certain amount of supersymmetry. These objects play a central role also in non--trivial tests of the AdS/CFT correspondence.

Supersymmetric Wilson loops have been first formulated in ${\cal N}=4$ SYM theory \cite{Maldacena, Rey:1998ik}, where adding a coupling to the scalars of the theory makes the Wilson loop operator locally invariant under half of the supercharges of the theory. In the context of the AdS/CFT correspondence, this 1/2 BPS object is the field theory dual of a fundamental macroscopic string living in the $AdS_5\times S^5$  background.
The expectation value of a 1/2 BPS Wilson loops on a path $\G$ can thus be computed at strong coupling in terms of a minimal area surface in $AdS_5$ ending on a contour $\G$ at its boundary and at a fixed point on $S^5$.

Whether or not supersymmetry is globally preserved depends on the shape of the contour.
For a straight line the normalized Wilson loops is invariant under eight supercharges and is protected in a way that its expectation value is exactly one.
For a circular contour the Wilson loop is still invariant under eight combinations of superconformal charges \cite{Bianchi:2002gz}. Despite being conformally equivalent to the Wilson loop on the straight line, it receives quantum corrections \cite{Erickson:2000af,Drukker:2000rr} due to the fact that the conformal transformation is anomalous. 
 
An exact result for the circular Wilson loop has been found in \cite{Pestun:2007rz} by applying localization techniques which allow to reduce its calculation to the evaluation of a finite dimensional matrix model. In the planar limit and for large coupling constant the matrix model result reproduces the string theory computation.
Therefore, it provides an example of function interpolating from weak to strong coupling, allowing for a very non--trivial test of the AdS/CFT correspondence.

A large class of less supersymmetric Wilson loops has also been constructed. In \cite{Drukker:2007dw,Drukker:2007qr} Wilson loops have been defined on a three sphere in space--time, which couple to scalars through the invariant one--forms of $S^3$ and preserve 1/16 of the supersymmetry.
When the Wilson loop contour is restricted to a maximal two sphere, supersymmetry gets enhanced and gives rise to 1/8 BPS configurations. Wilson loop expectation values can still be computed by a matrix model \cite{Drukker:2007yx,Pestun:2009nn}.
Amazingly, their expectation values depend only on the area enclosed by the contour on the two sphere and they turn out to be related to Wilson loops in two dimensional Yang--Mills theory.

In three dimensional Chern--Simons theory Wilson loops and their correlation functions are the basic observables to be computed and have been intensively studied in the past in connection with knot theory \cite{Witten}.

Renewed interest in three dimensional Wilson loops has grown after the formulation of the ABJM theory \cite{ABJM}, a three dimensional ${\cal N}=6$ superconformal Chern--Simons--matter model with gauge group $U(N)_k \times U(N)_{-k}$ (or $U(N)_k\times U(M)_{-k}$ in the ABJ generalization \cite{ABJ}) and opposite Chern--Simons levels $k$ and $-k$.
This theory describes the low energy dynamics of a stack on $N$ M2 branes in M--theory probing a $\mathbb{C}^4/{\mathbb{Z}_k}$ geometry.
In the large $N$ limit it has a 't Hooft parameter given by the ratio $\lambda = \frac{N}{k}$. For $k \gg N$ such a coupling is small and the theory allows for a perturbative description.
In the opposite regime the theory is strongly coupled and admits a dual description in terms of M--theory in the near horizon geometry of the M2 branes, that is $AdS_4 \times S^7/\mathbb{Z}_k$.
A supergravity approximation is valid when the radius of the M--theory circle is large, namely whenever $N\gg k^5$. In the intermediate region where $k\ll N\ll k^5$ the proper dual description is in terms of type IIA string theory on $AdS_4 \times \mathbb{CP}^3$.

The realization of a three dimensional version of the AdS/CFT correspondence has stimulated the study of observables in ABJM theory.
On the one hand scattering amplitudes of ${\cal N}=6$ supersymmetric Chern--Simons--matter theories \cite{ABL}--\cite{Bianchi:2013iha} have been computed and they have been found to exhibit nice symmetry properties as in four dimensional ${\cal N}=4$ SYM. 
Their allegedly dual objects, namely light--like polygonal Wilson loops have been also calculated \cite{HPW}--\cite{BGLP2}, and the emergence of a duality with amplitudes have been observed at four points.

On the other hand supersymmetric circular Wilson loop have been studied.
In particular, 1/6 BPS Wilson loops have been defined in \cite{Rey, Drukker, Chen}, which are formally similar to the 1/2 BPS Wilson loops of ${\cal N}=4$ SYM, in the sense that they only feature a coupling with the scalar fields of the theory \cite{Berenstein:2008dc}. Their expectation values have been computed in the planar limit, up to two loops.

As in ${\cal N}=4$ SYM, localization can be used to compute supersymmetric Wilson loops in ABJM.
The partition function of ABJM on a three-sphere has been shown to localize to a non--Gaussian matrix model \cite{Kapustin,MarinoPutrov}, and the expectation value of the 1/6 BPS Wilson loop has been computed by expanding the matrix model at weak coupling. In the planar limit the result is in agreement with the direct computation of \cite{Rey, Drukker, Chen}.
 
In parallel, a 1/2 BPS Wilson loop operator has been defined \cite{DrukkerTrancanelli}. Enhancing supersymmetry requires including a fermionic sector. As shown in \cite{DrukkerTrancanelli}, this is accomplished by extending the connection to a supermatrix of the $U(N|M)$ supergroup.
The 1/2 BPS Wilson loop has been proven to be cohomologically equivalent to a combination of 1/6 BPS Wilson loops. This has led to the possibility to use localization techniques to make a prediction \cite{DrukkerTrancanelli, Drukker:2010nc} for its expectation value using the already known results for 1/6 BPS objects.

Despite the result from localization, an explicit field theory perturbative computation of the two--loop contribution to the 1/2 BPS Wilson loop has been lacking until very recently.
In \cite{BGLP} we have tackled the problem of computing it in the planar limit, and comparing it with the weak coupling limit of the prediction coming from localization. 

The reconstruction of the localization result within an ordinary field theory approach is strongly motivated by many reasons. In fact, not only it represents a non--trivial check of the result, but it also allows for a deeper comprehension of the mechanisms underlying the cancellation of short distance divergences and the appearance of a finite result. Moreover, it addresses the question of understanding the relation between different regularization schemes. 

While in \cite{BGLP} we have simply reported the main result and briefly discussed its relation with the localization result, in the present paper we give a detailed explanation of the procedure we used and all technical aspects involved in the evaluation of contour integrals and discuss the role played by regulated short distance divergences. Moreover, we extend the previous result to the non--planar case, that is to finite $N,M$.  

The computation is hampered by intricate diagrams emerging from the fermionic sector.
The integrals appearing in the calculation are generally divergent and require regularization. We apply dimensional regularization with dimensional reduction (DRED).
The regulated integrals are hard to solve analytically, nevertheless we manage to perform them with the use of series expansions and Mellin--Barnes representation and   
provide results at any order in the regularization parameter. 

Within DRED scheme special care has to be taken in dealing with the ubiquitous $\varepsilon_{\mu\nu\rho}$ tensors of the Chern--Simons theory. We show how a consistent way to deal with them leads to the appearance of evanescent factors in front of divergent integrals. From the product of the two pieces finite terms arise, which concur to determine the final result. 

The comparison between the localization and the perturbative results requires a careful analysis of framing. In fact, while the perturbative calculation done using dimensional regularization corresponds to choosing framing zero, the localization result comes naturally in framing one. It is then necessary to identify the framing factor in the localization result and remove it. A proposal for the correct framing phase was made in \cite{DrukkerTrancanelli}. We have found that removing the factor proposed there, the remaining contribution perfectly matches the perturbative expression. Therefore, our perturbative result not only is a non--trivial check of the localization result but proves the correctness of the framing phase identified in \cite{DrukkerTrancanelli}. 
 
The main result of this paper is eq. (\ref{eq:result}) for the two--loop expansion of the 1/2 BPS Wilson loop for any value of $N,M$. The result is finite and perfectly matches the prediction from localization, color subleading terms included.

The paper is organized as follows. In Section \ref{sec:strategy} we give an upshot of the strategy we follow for the computation. In particular, we present the regularization we employ for taming divergent integrals, and outline the method for solving them.
In Section \ref{sec:oneloop} we compute the contributions to the Wilson loop expectation value at one loop and show that they are subleading in dimensional regularization.
In Section \ref{sec:twoloop} we present the diagrams contributing at two--loops.
These come from the purely bosonic sector and its mixing with fermions.
In Section \ref{sec:bosonic} we compute contributions from the former, which allows us to give a complete expression for the 1/6 BPS Wilson loop including subleading terms.
In Section \ref{sec:doublefermion} we solve the diagram coming from a double fermion exchange, while Section \ref{sec:gaugefermion} is devoted to the study of the intricate diagram featuring a mixed interaction vertex.
Some of these results require analytic continuation to be expanded in powers of the dimensional regularization parameter. This is accomplished in Section \ref{sec:expansions}.
Finally, in Section \ref{localization} we sum all contributions, so obtaining the 1/2 BPS Wilson loop expectation value at two loops and compare it to the prediction from localization, finding perfect agreement.
Several Appendices follow containing further details of the computation such as the explanation of the method for solving contour integrals and their explicit evaluation. \\

{\bf Note added:} A similar investigation has been performed in another paper \cite{GMPS}, cuncurrently appeared in the ArXiv, where the result of \cite{BGLP} has been reproduced using slightly different techniques for computing the integrals.

\section{The general strategy}
\label{sec:strategy}

We are interested in the perturbative evaluation of Wilson loops (WL) in $U(N)_k \times U(M)_{-k}$ ABJ(M) theories \cite{ABJM, ABJ}. These models contain two Chern--Simons gauge fields $A, \hat{A}$ plus propagating scalar matter fields $(C_I, \bar{C}^I)$, $I=1, ... , 4$,  and the corresponding fermions $(\bar{\psi}^I, \psi_I)$, all in the (anti)bifundamental representation of the gauge groups. The coupling to the gauge sector, as given by the action (\ref{action}), insures ${\cal N}=6$ supersymmetry. We work in euclidean three--dimensional space. Conventions and definitions are collected in Appendix \ref{sec:conventions}. 

WL preserving $1/6$ of the supersymmetry have been constructed as \cite{Rey,Drukker,Chen}
\bea
&& \langle W_{1/6}[\G] \rangle =  \frac{1}{N} \int  D[A, \hat{A}, C, \bar{C}, \psi, \bar{\psi}] \; e^{-S}  \; \Tr \left[ P \exp{
\left( - i \int_\G d\tau {\cal A}(\tau) \right) } \right]
\non \\
&& \langle \hat{W}_{1/6}[\G] \rangle =  \frac{1}{M} \int  D[A, \hat{A}, C, \bar{C}, \psi, \bar{\psi}] \; e^{-S}  \; \Tr \left[ P \exp{
\left( - i \int_\G d\tau {\hat{\cal A}}(\tau) \right) } \right]
\eea
where $S$ is the euclidean action (\ref{action}), $\G$ is an infinite straight line or, equivalently, a circle and the generalized connections are defined as
\beq
\label{connections}
 {\cal A} = A_\mu \dot{x}^\mu - \frac{2\pi i}{k}  |\dot{x}| {\mathcal M}_J^{\; I} C_I \bar{C}^J
\qquad , \qquad 
\hat{\cal A} = \hat{A}_\mu 
\dot{x}^\mu - \frac{2\pi i}{k}  |\dot{x}| \hat{{\mathcal M}}_J^{\; I}   \bar{C}^J C_I
\eeq
in terms of two constant matrices ${\mathcal M} = \hat {\mathcal M}={\rm diag}(-1,-1,1,1)$.

A linear combination of these WL transforming oppositely under time--reversal
is the $1/6$ BPS operators studied in Ref$.$ \cite{Rey}, that is
\begin{equation}
\label{eq:Wplus}
W_{1/6}^{+}[\G] = \frac{N\, W_{1/6}[\G] + M\, \hat{W}_{1/6}[\G]}{N+M}
\end{equation}

As discussed in \cite{Drukker}, a suitable extension of the previous holonomies allows for the construction of a WL that preserves 1/2 of the supersymmetry.
This can be expressed in terms of a superconnection ${\cal L}$ of the supergroup $U(N|M)$. Precisely, the 1/2 BPS WL is defined as
\bea
\label{WL}
\langle W_{1/2}[\G] \rangle = \frac{1}{M+N} \int D[A, \hat{A}, C, \bar{C}, \psi, \bar{\psi}] \; e^{-S}  \; \Tr \left[ P \exp{
\left( - i \int_\G d\tau {\cal L}(\tau) \right) } \right]
\eea
where ${\cal L}(\tau)$ is represented as the supermatrix  
\beq
\label{supermatrix}
{\cal L}(\tau) \, = \, \left( \begin{array}{ccc}
{\cal A} & -i \sqrt{\frac{2\pi}{k}} |\dot{x}| \eta_I \bar{\psi}^I \\
 -i \sqrt{\frac{2\pi}{k}}  |\dot{x}| \psi_I \bar{\eta}^I & \hat{\cal A}  \end{array}\right)   
\eeq
with ${\cal A}, \hat{\cal A}$ still given by (\ref{connections}), though with different matrices ${\cal M}, \hat{{\cal M}}$ (see (\ref{DT}) below) and
$\eta, \bar{\eta}$ are two commuting spinors controlling the couplings to the fermions in the (anti)bifundamental representation of the gauge groups. 

We choose the path $\G$ to be the unit circle, parametrized as 
\beq
\label{circle}
x^\mu = (0, \cos{\tau}, \sin{\tau}) \quad , \quad \tau \in [0, 2\pi]
\eeq
In this case, preserving half of the supersymmetry requires \cite{DrukkerTrancanelli}
\bea
\label{DT}
&& {\mathcal M}_I^{\; J} = \hat{{\mathcal M}}_I^{\; J} = \d^J_I - 2 \d^J_1 \d_I^1 ,
\\
&& \eta_I^\a (\tau) = \left( e^{i\tau/2} \quad -i  e^{-i\tau/2} \right) \, \d^1_I  \quad , \quad \bar{\eta}_\a^I(\tau) = \left( \begin{array}{c}  
i  e^{-i\tau/2} \\
- e^{i\tau/2} 
 \end{array}\right) \, \d_1^I \quad , \qquad (\eta \bar{\eta}) = 2i .
 \non
\eea

The perturbative evaluation of the expression (\ref{WL}) is performed by Taylor expanding the exponential of the superconnection and taking the expectation value by Wick contracting the fields. Since we are interested in the two--loop quantum corrections, it suffices to expand it up to the fourth order. In this process we get purely bosonic contributions from the diagonal part of the $U(N|M)$ super-matrix (\ref{supermatrix}), purely fermionic contributions from the off--diagonal blocks and mixed contributions from the mixing of the two.
We will study the bosonic and fermionic contributions separately. 

When computing loop integrals and performing integrations along the circle, potential divergent contributions arise at short distances. 
In order to deal with them, we use dimensional regularization with dimensional reduction scheme (DRED) \cite{Siegel}, which has been proven \cite{Chen:1992ee}  
to preserve gauge invariance and supersymmetry of Chern--Simons theories up to two loops \footnote{Recently, this prescription has been also shown to fix the problem of uniform (maximal) transcendentality of the two--loop result for the light--like tetragon WL in ABJM theory \cite{BGLP2}.}. 

DRED requires to assign Feynman rules in three dimensions and perform all tensor manipulations strictly in three dimensions before promoting loop integrals to $D= 3-2\epsilon$. A suitable prescription is then required for contracting three--dimensional objects coming from Feynman rules with $D$--dimensional tensors arising from tensorial integrals.
DRED scheme assigns the following rules  \cite{Siegel:1980qs} for contracting three--dimensional metrics $\eta^{\mu\nu}$ and $D$--dimensional ones $\hat \eta^{\mu\nu}$ 
\beq
\label{DRED}
\eta^{\mu\nu} \eta_{\mu\nu} =3 \qquad \quad 
\hat{\eta}^{\mu\nu} \hat{\eta}_{\mu\nu} = 3-2\e \qquad \quad 
\eta^{\mu\nu} \hat{\eta}_{\nu\rho} = \hat \eta^{\mu}_{\phantom{\mu}\rho}
\eeq
Particular care is required when contracting $D$--dimensional metric tensors with Levi--Civita tensors $\varepsilon_{\mu\nu\rho}$ that cannot be defined outside three dimensions.  Usually, two possible strategies for overcoming the problem can be used: Either tensor algebra is performed until one reaches a situation where only scalar integrals survive \cite{Chen:1992ee}, or one applies algebraic identities in order to get rid of all $\varepsilon$ tensors. In the following, we will adopt the latter strategy; in particular, products of  $\varepsilon$ tensors will be traded with products of three--dimensional metric tensors via the identity
\beq
\label{epsilon}
\varepsilon_{\lambda\mu\nu} \varepsilon_{\rho\sigma\tau} = \eta_{\l \rho} ( \eta_{\mu\s} \eta_{\nu\tau} - \eta_{\mu\tau} \eta_{\nu\s} ) 
- \eta_{\l \s} ( \eta_{\mu\rho} \eta_{\nu\tau} - \eta_{\mu\tau} \eta_{\nu\rho} ) + \eta_{\l \tau} ( \eta_{\mu\rho} \eta_{\nu\s} - \eta_{\mu\s} \eta_{\nu\rho} ) 
\eeq

When parametrizing the WL circular contour by polar coordinates, the final integrals take the form of multiple integrations over powers of trigonometric functions.
In particular, such powers depend on the regularization parameter $\epsilon$ and therefore it is necessary to carry out such integrals analytically for all values of $\epsilon$.
This is a hard task which we tackle as explained in detail in Appendix \ref{sec:method}.
The central idea of our method is the following. After rewriting trigonometric functions as complex exponentials, we expand powers as geometric series. Working in save regions of the parameters where we can exchange series with integrals we can then easily perform integrations term by term, producing multiple series. Finally, we manage to sum such series in terms of hypergeometric functions.
 
In general, once regularized, the integrals converge in the complex half--plane defined by some critical value of the real part of the regularization parameter $\epsilon$.
In the region of parameters where these functions are well--defined we test the results numerically.

In general, the half--plane where integrals and hypergeometric series converge does not include the neighbourhood of $\epsilon=0$, which is the physical limit we need eventually take. Therefore, we analytically continue the hypergeometric functions close to the $\epsilon=0$ region and expand the results up to finite terms.

\section{WL at one loop}
\label{sec:oneloop}

At one loop, there are in principle three diagrams. Two of them come from the purely bosonic sector and are also present in the computation of the 1/6 BPS WL.
These are a scalar tadpole arising from the first order expansion of the exponential and a single gluon exchange, from expanding the WL operator to second order.
\FIGURE{
\centering
 \includegraphics[width=0.2\textwidth]{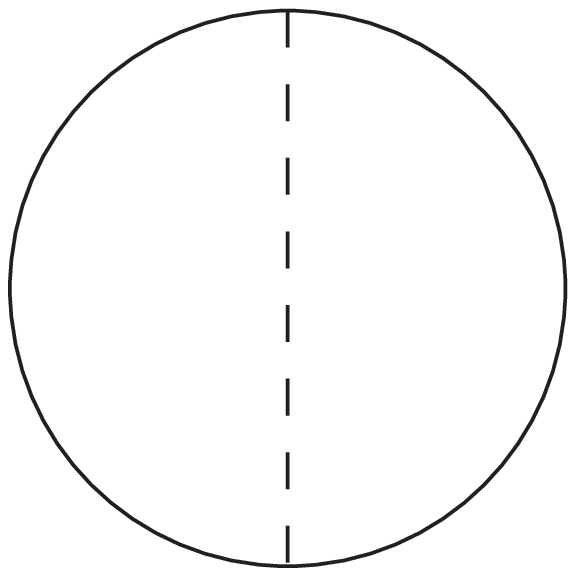}
    \caption{One--loop fermion exchange diagram.}
    \label{1loop}
}
Since we work in dimensional regularization we consistently discard the tadpole diagram.
The gluon exchange vanishes because of the antisymmetry of the $\varepsilon$ tensor carried by the propagator (\ref{treevector}), which is contracted with three vectors lying on the plane of the circular contour. 

We stress that this feature is special of performing the computation without framing \cite{Witten,GMM}. The inclusion of framing, as reviewed in Section \ref{localization}, consists in thickening the WL contour with an infinitesimally displaced framing path, which may wound around the original one.
This provides a point--splitting regularization of the WL.
In such an approach the gluon exchange diagrams should have been taken into account, since, for instance, the $\varepsilon$ tensor could be contracted with a vector slightly off the plane where the circle lies.
This kind of simplification from working without framing occurs also at two--loops.

The extra diagram comes from the fermionic sector and contributes to the 1/2 BPS WL only. This is a single fermion exchange depicted in Fig. \ref{1loop}, from the second order expansion of the WL operator. Contrary to the previous diagrams this does not vanish by symmetry and is the only non--trivial contribution to the WL.

Inserting the explicit expression for the fermion propagator (\ref{treefermion}) and forgetting about overall coefficients we obtain
\begin{align}
\langle W_{1/2}[\G] \rangle^{(1)}  &\sim  \int d\tau_{1>2} \; \eta_I^\a(\tau_1) \, \langle \bar{\psi}_\a^I (\tau_1) \, \psi_J^\b (\tau_2) \rangle \, \bar{\eta}^J_\b (\tau_2) 
\nonumber\\&
\sim   \int d\tau_{1>2} \; (\eta_1 \g^\mu \bar{\eta}_2) \, \frac{(x_1 - x_2)_\mu}{[(x_1 - x_2)^2]^{\frac32-\e}} 
\end{align}
where we used the notation $\int d\tau_{1>2} \equiv \int_0^{2\pi} d\tau_1 \int_0^{\tau_1} d\tau_2$. The integral is reduced to 
\beq
\label{IntegralW12}
\langle W_{1/2}[\G] \rangle^{(1)} \sim  I^{(1)} = \int_0^{2\pi} d\tau_1 \int_0^{\tau_1} d\tau_2  \; \frac{1}{[\sin^2{\frac{\tau_{12}}{2}}]^{1-\e}} ,
\eeq
as it can be easily verified by using identities (\ref{I1}, \ref{eq:etagammax}).

The evaluation of integral (\ref{IntegralW12}) is discussed in greater detail in Appendix \ref{sec:method}; here we simply give an upshot of the procedure. 

After writing the trigonometric function in terms of exponentials, we expand powers as geometric series and perform integrations term by term.  Imposing the result to be real 
for $\epsilon$ real allows to express the integral as
\beq
I^{(1)} = \frac{2^{3-2\epsilon}\, \pi }{\Gamma(2-2\epsilon)}
\sin(\pi(1-\epsilon)) S_1[1-\epsilon]
\eeq
where the series $S_{\l}[\a]$ is defined in (\ref{Sseries}). This series can be summed, see (\ref{eq:oneloopseries}), yielding
\begin{equation}
I^{(1)} = \frac{2 \pi ^{3/2} \Gamma \left(-\frac{1}{2}+\epsilon \right)}{\Gamma \left(\epsilon \right)}
\end{equation}
The integral turns out to be subleading in $\epsilon$. Therefore, removing the regularization parameter we get
\beq
\label{WLoneloop}
\left\langle W_{1/2}[\G]\right\rangle^{(1)}  =0 ,
\eeq
in line with the prediction from localization, as we discuss in Section \ref{localization}.

\section{WL at two loops: The diagrams}
\label{sec:twoloop}

At two loops, neglecting contributions which vanish identically because of the antisymmetry of the $\varepsilon$ tensor, we are left with the diagrams in Fig. \ref{2loop}. We note that, due to the bifundamental nature of matter, non--planar diagrams cannot be constructed from scalars and fermions.  In the gauge sector, one potential non--planar contribution would be a crossed ladder pure gauge graph, but it turns out to be identically zero like the planar ladder. Therefore, subleading contributions come only from the pure gauge graph 2(a).

The first three diagrams in Fig. \ref{2loop} are purely bosonic diagrams coming from contracting the diagonal terms in (\ref{supermatrix}). Apart from the difference in the matrices ${\cal M},\hat{{\cal M}}$ these contributions are common to 1/2 BPS and 1/6 BPS WL. What distinguishes the 1/2 BPS WL is the appearance of three extra contributions with fermion propagators corresponding to contractions which involve off--diagonal terms in the supermatrix ${\cal L}$.  

\FIGURE{
\centering
 \includegraphics[width=0.75\textwidth]{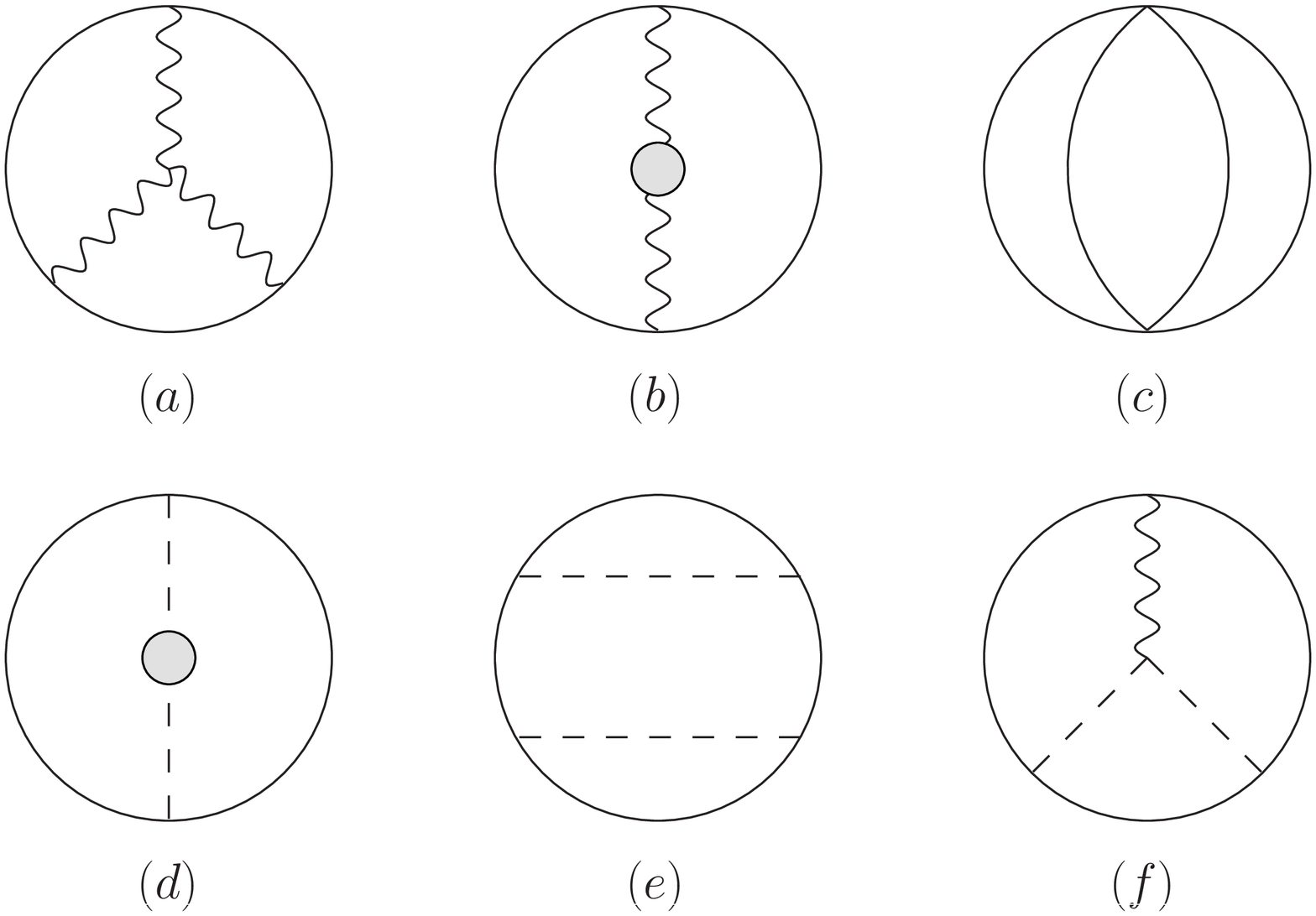}
    \caption{Non-vanishing two--loop diagrams for BPS Wilson loops. Wavy lines represent gauge propagators, solid lines represent scalars, and dashed lines are fermion propagators. Bubbles represent one--loop corrections to the propagators. }
    \label{2loop}
}

In particular, diagram 2(d) comes from the fermionic part of the second order expansion of the WL with a one--loop contraction between the two fermions. Using the fermion propagator 
(\ref{1fermion}) we obtain
\beq
{\rm (d)} = i \frac{M-N}{M+N} \, \left( \frac{MN}{k^2} \right) \frac{\G^2(\frac12 -\e)}{4\pi^{1-2\e}} \, \int d\tau_{1>2} \, 
\frac{|\dot{x}_1| |\dot{x}_2|}{[(x_1-x_2)^2]^{1-2\e}} \, \left[ (\eta_1 \bar{\eta}_2) - (\eta_2 \bar{\eta}_1) \right] = 0
\eeq
The last equality easily follows from the fact that $(\eta_1 \bar{\eta}_2) = (\eta_2 \bar{\eta}_1)$, as is evident from identity (\ref{eq:etaetabar}). Therefore, diagram 2(d) does not contribute. 

We are then left with three bosonic contributions plus two non--vanishing fermionic contributions corresponding to a double fermion--line diagram, Fig. \ref{2loop}(e), and a gauge--fermion vertex diagram, Fig. \ref{2loop}(f). We are going to discuss them separately. \\

\section{WL at two loops: The bosonic sector}
\label{sec:bosonic}

The first three diagrams in Fig. \ref{2loop} have been already evaluated in literature \cite{Rey} for the 1/6 BPS case, in the planar limit. Here we briefly review the results, extending them to the case of $M$, $N$ finite. 

Diagram 2(a) comes from the gauge part of the third order expansion of the WL contracted with the gauge cubic vertex (\ref{gaugecubic}). Summing the contributions from the two generalized connections ${\cal A}$ and $\hat{\cal A}$, we can write 
\beq
\label{int}
{\rm (a)} =  - \frac{M(M^2-1) + N(N^2-1)}{M+N} \, \frac{1}{k^2} \, 
\frac{\G^3(\frac32 - \e)}{2\pi^{\frac{5}{2} - 3\e}}
\int d\tau_{1>2>3} \, \dot{x}_1^\s \, \dot{x}_2^\eta \, \dot{x}_3^\zeta \, \varepsilon^{\xi \tau \kappa} \varepsilon_{\s\xi\mu} \varepsilon_{\eta\tau\nu} \varepsilon_{\zeta \kappa \rho} 
\,  I^{\mu\nu\rho}
\eeq
where 
\beq
\label{intI}
I^{\mu\nu\rho} \equiv \int d^{3-2\e} x  \,  
\frac{  (x-x_1)^\mu (x-x_2)^\nu (x-x_3)^\rho}{|x-x_1|^{3-2\e} |x-x_2|^{3-2\e} |x-x_3|^{3-2\e} } 
\eeq
Integral (\ref{int}) is well--known from pure Chern--Simons and, being finite, can be computed at $\e =0$ (see for instance ref. \cite{Rey}, eq. (6.11)). Its value is $ \frac{8}{3} \pi^3$, so we obtain
\beq
{\rm (a)} =   - \frac{M^2 + N^2 - MN - 1}{k^2}  \, \frac{\pi^2}{6}
\eeq
 
Diagrams 2(b) and 2(c) arise from the bosonic part of the quadratic term in the WL expansion where in one case the one--loop gauge propagator is inserted, whereas in the second case contractions in the scalar sector are performed. 

Using the 1--loop vector propagator (\ref{1vector}), the first term gives
\beq
{\rm (b)} = - \frac{NM^2+ N^2M}{M+N} \, \frac{1}{k^2} \, \frac{\G^2(\frac12-\e)}{\pi^{1 -2\e}} \; \int d\tau_{1>2} \, \frac{\dot{x}_1 \cdot \dot{x}_2}{[(x_1-x_2)^2]^{1-2\e}}
\eeq
whereas, using the scalar propagator (\ref{scalar}) the second term gives
\beq
{\rm (c)} =  \frac{NM^2+ N^2M}{M+N} \, \frac{1}{k^2} \, \frac{\G^2(\frac12-\e)}{4\pi^{1 -2\e}} \; \int d\tau_{1>2} \, \frac{|\dot{x}_1| |\dot{x}_2|}{[(x_1-x_2)^2]^{1-2\e}}
\, \Tr({\mathcal M}_1{\mathcal M}_2)
\eeq
Summing the two contributions and using identity (\ref{eq:matrices}) we obtain
\beq
\label{bc}
[{\rm (b) + (c)}] =  \frac{MN}{k^2} \, \frac{\G^2(\frac12-\e)}{\pi^{1 -2\e}} \; \int d\tau_{1>2} \, \frac{- \dot{x}_1 \cdot \dot{x}_2 + |\dot{x}_1| |\dot{x}_2|}{[(x_1-x_2)^2]^{1-2\e}}
\eeq
The integral is finite and can be computed exactly in three dimensions. For the unit circle it gives $\pi^2$ (see eq. (6.10) in \cite{Rey}) and we obtain 
\beq
[{\rm (b) + (c)}] =   \frac{MN}{k^2} \, \pi^2
\eeq

\noindent
Summing diagrams 2(a), 2(b) and 2(c) from the bosonic sector and normalizing properly gives rise to the expectation value of the 1/6 BPS WL combination at two loops
\begin{equation}
{\rm (a) + (b) + (c)} = \langle W_{1/6}^{+} \rangle^{(2)}
\end{equation}
The result is

\beq\label{eq:bosonicresult}
\boxed{\boxed{\langle W_{1/6}^{+}  \rangle^{(2)} = 1 + \frac{\pi^2}{6\, k^2} \left[ -(M^2 + N^2) + 7 MN + 1 \right] } }
\eeq

We note that for $M=N$ and in the planar limit the two--loop correction reduces to $\tfrac{5}{6} \pi^2 \left( \tfrac{N}{k} \right)^2$, which is the result for the 1/6 BPS WL in ABJM \cite{Rey}. \\

\section{WL at two loops: Double fermion--line diagram}
\label{sec:doublefermion}

We now move to the evaluation of new genuine contributions to the 1/2 BPS WL.  
 
Diagram 2(e) corresponds to a double contraction of fermions in the fourth order expansion of the WL. Using the tree--level fermionic propagator (\ref{treefermion}) and summing over the two possible ways to make contractions, we obtain
\begin{align}
{\rm (e)} &= - \frac{1}{M+N} \left( \frac{2\pi}{k} \right)^2 \frac{\G^2(\frac32 -\e)}{4 \pi^{3-2\e}} \, \int d\tau_{1>2>3>4} \,  |\dot{x}_1| |\dot{x}_2| |\dot{x}_3| |\dot{x}_4| \times 
\\
& \left\{ \left[ MN^2 (\eta_1 \g^\mu \bar{\eta}_2) (\eta_3 \g^\nu \bar{\eta}_4)  + M^2 N (\eta_2 \g^\mu \bar{\eta}_1) (\eta_4 \g^\nu \bar{\eta}_3) 
\right] \, \frac{ (x_1 - x_2)_\mu (x_3 - x_4)_\nu}{[ (x_1-x_2)^2 (x_3-x_4)^2]^{\frac32 -\e}} \right.
\non \\
& \left. \quad +  \left[ M^2 N  (\eta_1 \g^\mu \bar{\eta}_4) (\eta_3 \g^\nu \bar{\eta}_2)  + M N^2 (\eta_4 \g^\mu \bar{\eta}_1) (\eta_2 \g^\nu \bar{\eta}_3) \right] \, \frac{ (x_1 - x_4)_\mu (x_2 - x_3)_\nu}{[ (x_1-x_4)^2 (x_2-x_3)^2]^{\frac32 -\e}} \right\} \non 
\end{align}
This expression can be easily elaborated by the help of identity (\ref{eq:etagammax}) and we find
\begin{equation}
\label{eresult}
{\rm (e)} = \frac{MN}{k^2} \, \frac{\G^2(\frac32 -\e)}{(4\pi)^{1-2\e}} \, \int d\tau_{1>2>3>4} \;
\left\{ \frac{1}{[\sin^2{\frac{\tau_{12}}{2}} \sin^2{\frac{\tau_{34}}{2}}]^{1 -\e}} + \frac{1}{[\sin^2{\frac{\tau_{14}}{2}} \sin^2{\frac{\tau_{23}}{2}}]^{1-\e}}  \right\}
\end{equation}
where we have introduced the notation $\tau_{ij} = \tau_i - \tau_j$.  

Neglecting for a moment the overall coefficient in front, we concentrate on the evaluation of the two integrals
\beq
\label{planarintegrals}
I_{\rm e}^{(1)}+I_{\rm{e}}^{(2)}= \int\limits_{0}^{\,\,2\pi} d\t_1\int\limits_{0}^{\,\,\t_1}d\t_2
\int\limits_{0}^{\,\,\t_2}d\t_3 \int\limits_{0}^{\,\,\t_3}d\t_4
\left\{ \frac{1}{[\sin^2{\frac{\tau_{12}}{2}} \sin^2{\frac{\tau_{34}}{2}}]^{1 -\e}} + \frac{1}{[\sin^2{\frac{\tau_{14}}{2}} \sin^2{\frac{\tau_{23}}{2}}]^{1-\e}}  \right\}
\eeq
The easiest way to carry out the calculation is to trade the sum of these two integrals with a third one  
\beq
\label{crossedintegral}
I_{\rm e}^{(3)}\equiv 
\int\limits_{0}^{\,\,2\pi} d\t_1\int\limits_{0}^{\,\,\t_1}d\t_2
\int\limits_{0}^{\,\,\t_2}d\t_3 \int\limits_{0}^{\,\,\t_3}d\t_4
\frac{1}{[\sin^{2} \frac{\t_{13}}{2} \sin^{2} \frac{\t_{24}}{2}]^{1-\e}}
\eeq
which would correspond to a crossed diagram. This is based on the observation that the sum $(I_{\rm e}^{(1)}+I_{\rm e}^{(2)}+I_{\rm e}^{(3)})$ is subleading in $\e$, as shown at the end of Appendix \ref{sec:method}. Therefore, up to ${\cal O}(\e)$ terms, we can write $(I_{\rm e}^{(1)}+I_{\rm e}^{(2)}) = -I_{\rm e}^{(3)}$. 

The evaluation of  $I_{\rm e}^{(3)}$ turns out to be much simpler, being it finite for $\e \rightarrow 0$.  We compute it by generalizing to the two--loop case the method outlined in Appendix \ref{sec:method}. The main difference in this case is that we have to expand the product of two trigonometric functions. As a result, performing the integrations term by term, we end up with a linear combination of double series. Precisely, if we introduce the notation
\beq
S_{\lambda_1,\lambda_2,\lambda_3}[\alpha]=
\sum\limits_{n,m=0}^{\infty}\frac{\Gamma(n+2\alpha)\Gamma(m+2\alpha)}
{n!m!(n+\alpha)^{\lambda_1}(m+\alpha)^{\lambda_2}(m+n+2\alpha)^{\lambda_3}}
\eeq
and define $(1 - \e) \equiv \a$ for convenience, then imposing that the result of the integral be real for $\a$ real as explained in Appendix \ref{sec:method}, we find
\bea
\label{error3}
I_{\rm e}^{(3)} &=& 
\frac{2^{4 \alpha }}{\Gamma^2 (2 \alpha )}
\Big[ 
- 2 \pi  \sin (2 \pi  \alpha ) \, S_{1,1,1}[\a] + \left(1-\cos (2 \pi  \alpha )\right)\, S_{2,2,0}[\a]
\Big] \\
&=& 
\frac{2^{4 \alpha }}{\Gamma^2(2 \alpha )}
\Big[ 
- 2 \pi  \sin (2 \pi  \alpha ) \, \left( S_{2,1,0}[\a] - S_{2,0,1}[\a] \right) + \left(1-\cos (2 \pi  \alpha )\right)\, S_{2,2,0}[\a]
\Big] \non 
\eea
where in the second line a simple algebraic identity has been used.

Numerical evaluation over a large range of complex values of $\alpha$ shows that this result is correct. 
In order to obtain an analytical result we have to evaluate the previous series. Using known summation formulae we obtain
\bea
S_{2,1,0}[\a] &=& 2^{-4 \alpha } \cot (\pi  \alpha ) \Gamma^2 \left(\frac{1}{2}-\alpha \right) \Gamma^2 (\alpha ) \Gamma^2 (2 \alpha )
\non \\
S_{2,0,1}[\a] &=& \frac{\pi \Gamma^2 (2 \alpha ) \, _4F_3(\alpha ,\alpha ,2 \alpha ,2 \alpha ;1,\alpha +1,\alpha +1;1)}{\alpha ^2\, \sin (2 \pi  \alpha )}
\eea
and 
\beq
S_{2,2,0}[\a] = \pi  2^{-4\alpha } \cot ^2(\pi  \alpha ) \Gamma^2 \left(\frac{1}{2}-\alpha \right) \Gamma^2 (\alpha ) \Gamma^2 (2 \alpha )
\eeq
Collecting all pieces and taking into account the overall coefficient in eq. (\ref{eresult}) we finally obtain 
\bea
\label{eq:I2f}
{\rm (e)} &=& \frac{MN}{k^2} \, \Big\{  - 8 \pi^{1+2\e} \frac{\G^2\left( \frac32 -\e \right)}{(1-\e)^2} \,  _4F_3\left[
             \begin{array}{c}
             2\!-\!2\epsilon,2\!-\!2\epsilon,1\!-\!\epsilon,1\!-\!\epsilon \\
             1,2\!-\!\epsilon,2\!-\!\epsilon \\
             \end{array};1
             \right]
\non \\             
&~& +\frac12 \, (2\pi)^{2\e} \,   \Gamma^2 (1-\epsilon ) \, \Gamma^2 
\left(\frac{1}{2} -\e \right)  \, \Gamma^2 \left(\frac{1}{2} +\e \right)
\, \cos^2(\pi\epsilon) \Big\} 
\eea
 
This expression is well--defined in the complex half--plane $\Re(\e)>\tfrac{1}{4}$. Therefore, an analytical continuation of the result in a region close to $\e=0$ is required in order to obtain a safe expansion in powers of $\e$. We postpone the discussion of this point to Section \ref{sec:expansions}. \\

\section{WL at two loops: Gauge--fermion vertex diagram}
\label{sec:gaugefermion}

Diagram 2(f) arises from the six gauge--fermion mixed terms in the third order expansion of the WL (recall that $|\dot{x}| =1$ on the unit circle)
\begin{align}
\label{3expansion}
- \frac{i}{M+N} \left( \frac{2\pi}{k} \right) \, \int d\tau_{1>2>3} \,  
& \Tr 
\Big\{  \eta_{2  I} \bar{\eta}_3^J  \, \langle A_{1\mu} \bar{\psi}_2^I \psi_{3  J} \rangle \, \dot{x}_1^\mu  
~+~ \bar{\eta}_2^I \eta_{3 J}  \, \langle \hat{A}_{1\mu} \psi_{2 I} \bar{\psi}_3^J \rangle \, \dot{x}_1^\mu  
\\
&~~ 
+ \eta_{3 I} \bar{\eta}_1^J  \, \langle  \psi_{1  I} A_{2 \mu} \bar{\psi}_3^J \rangle \, \dot{x}_2^\mu  
~+~ \bar{\eta}_3^I \eta_{1 J}  \, \langle \bar{\psi}_1^J \hat{A}_{2\mu} \psi_{3 I}   \rangle \, \dot{x}_2^\mu  
\non \\
&~~
+ \eta_{1 I} \bar{\eta}_2^J  \, \langle   \bar{\psi}_1^I \psi_{2 J} A_{3 \mu} \rangle \, \dot{x}_3^\mu \,  
~+~ \bar{\eta}_1^I \eta_{2 J}  \, \langle \psi_{1  I} \bar{\psi}_2^J \hat{A}_{3\mu} \rangle \, \dot{x}_3^\mu \,  \Big\} 
\non
\end{align}
contracted with one mixed vertex (\ref{gaugefermion}) coming from the action. 
 
Computing for instance the first term we obtain
\beq
\label{first}
I_{\rm f}^{(1)} = - \frac{N}{M+N} \left( \frac{M}{k} \right)^2 \frac{\G^3(\frac12 -\e)}{16 \pi^{\frac52 - 3\e}} \int d\tau_{1>2>3}  
( \eta_2 \g^\xi \g^\mu \g^\s \bar{\eta}_3) \, \dot{x}_1^\nu \, \varepsilon_{\nu \mu}^{\phantom{\nu \mu} \rho} \, \G_{\rho \xi \sigma}
 \eeq
where we have defined (see eq. (\ref{intI}))
\bea
\label{gammaintegrals}
\Gamma^{\mu\nu\rho} &\equiv& ( 1 - 2\e )^3 \,  I^{\mu\nu\rho} \, = \,
\partial^{\mu}_1\, \partial^{\nu}_2\, \partial^{\rho}_3\, \int \frac{d^3 x}{[ (x-x_1)^2 (x-x_2)^2 (x-x_3)^2]^{\frac12 -\e}} 
\eea
The integral $I_{\rm f}^{(2)} $ for the second term in (\ref{3expansion}) can be easily obtained from $I_{\rm f}^{(1)} $ by exchanging $M \leftrightarrow N$, $2  \leftrightarrow 3$ and multiplying by $(-1)$. Analogously, the sum of the third and fourth terms and the sum of the fifth and the sixth terms can be easily obtained from $I_{\rm f}^{(1)} + I_{\rm f}^{(2)}$ by permuting the indices $1,2,3$ (the second term picks up a minus sign from the exchange of the two fermions). Therefore, it is sufficient to concentrate on (\ref{first}). The spinorial structure appearing there can be simplified by using the identity in (\ref{id1}) for the product of three gamma matrices. As a consequence, the contribution (\ref{first}) takes the form
\bea
\label{1fintegral}
&& I_{\rm f}^{(1)}  = - \frac{N}{M+N} \left( \frac{M}{k} \right)^2 \frac{\G^3(\frac12 -\e)}{16 \pi^{\frac52 - 3\e}} \int d\tau_{1>2>3} 
\times
\Big\{ (\eta_2 \g_\mu \bar{\eta}_3) \varepsilon_{\nu \rho \s} \, \dot{x}_1^\nu \Big(  \Gamma^{\s\rho\mu} + \Gamma^{\s\mu\rho}   \Big)
\non \\
\non \\
&& \qquad\qquad - i (\eta_2 \bar{\eta}_3) \dot{x}_1^{\nu} \, \left( \Gamma^{\mu}_{\phantom{\mu}\nu\mu}  - \Gamma^{\mu}_{\phantom{\mu}\mu\nu} \right) 
-  (\eta_2 \g^0 \bar{\eta}_3) \varepsilon_{\rho 0 \nu} \, \dot{x}_1^\rho \Gamma^{\nu\mu}_{\phantom{\nu\mu}\mu}  \Big\}
\eea
where in the third piece we have taken into account that for the planarity of the contour non--vanishing contributions arise only when one of the indices in the $\varepsilon$ tensor is zero.
Summing the $I_{\rm f}^{(2)}$ piece amounts to adding a contribution $-I_{\rm f}^{(1)} \big|_{2\leftrightarrow 3, M \leftrightarrow N}$.
Exploiting the symmetry properties of $\eta(\t_i) \bar\eta(\t_j)$, $\eta(\t_i) \g^0 \bar\eta(\t_j)$ under the exchange $i \leftrightarrow j$, as follows from identities (\ref{eq:etaetabar}, \ref{eq:etagamma0}) and observing that $\G^\nu_{\phantom{\nu} \mu \nu} (k,i,j) =  \G^\nu_{\phantom{\nu} \nu \mu} (k,j,i)$, it can be rewritten as 
$ I_{\rm f}^{(2)} = I_{\rm f}^{(1)}\big|_{M\leftrightarrow N}$. Therefore, combining the color factors, we finally obtain
\bea
\label{12fintegral}
&& I_{\rm f}^{(1)} + I_{\rm f}^{(2)} =  - \frac{MN}{k^2}\, \frac{\G^3(\frac12 -\e)}{16 \pi^{\frac52 - 3\e}} \int d\tau_{1>2>3} 
\times
\Big\{ (\eta_2 \g_\mu \bar{\eta}_3) \varepsilon_{\nu \rho \s} \, \dot{x}_1^\nu \Big(  \Gamma^{\s\rho\mu} + \Gamma^{\s\mu\rho} \Big)
\non \\
\non \\
&& \qquad\qquad 
- i (\eta_2 \bar{\eta}_3) \dot{x}_1^{\nu} \, \left( \Gamma^{\mu}_{\phantom{\mu}\nu\mu}  - \Gamma^{\mu}_{\phantom{\mu}\mu\nu} \right) 
-  (\eta_2 \g^0 \bar{\eta}_3) \varepsilon_{\rho 0 \nu} \, \dot{x}_1^\rho \Gamma^{\nu\mu}_{\phantom{\nu\mu}\mu}  \Big\}
\eea

The same pairing occurs for $I_{\rm f}^{(3)} + I_{\rm f}^{(4)}$ and $I_{\rm f}^{(5)}+I_{\rm f}^{(6)}$, whose results can be easily obtained from the previous one by permuting the indices 1,2,3 (with care to the signs, as already mentioned). 

The expression (\ref{12fintegral}) is particularly complicated to compute. In order to simplify the procedure it is convenient to separate it 
into the sum of two pieces, one proportional to $\Gamma ^{\mu \nu \rho }$ integrals (\ref{gammaintegrals}) with indices contracted with external tensor structures (uncontracted integrals, denoted $U_{\rm f}$) and one proportional to $\Gamma ^{\mu \nu \rho }$ integrals with a pair of contracted indices (contracted integrals, denoted $C_{\rm f}$). We then write
\beq
\label{CU}
I_{\rm f}^{(1)} + I_{\rm f}^{(2)} + I_{\rm f}^{(3)} + I_{\rm f}^{(4)} + I_{\rm f}^{(5)} + I_{\rm f}^{(6)} \equiv 
  - \frac{MN}{k^2}\, \frac{\G^3(\frac12 -\e)}{16 \pi^{\frac52 - 3\e}} \, \Big( U_{\rm f} + C_{\rm f} \Big)
\eeq
We devote the next two Sections to the hard task of evaluating $U_{\rm f}$ and $C_{\rm f}$.

\subsection{The uncontracted integrals $U_{\rm f}$}
\label{sec:uncontracted}
 
We first concentrate on the evaluation of the part of diagram \ref{2loop}(f) corresponding to $\Gamma ^{\mu \nu \rho }$ integrals with indices contracted with external tensor structures, namely the contributions denoted by $U_{\rm f}$. Precisely, summing over all contributions we need evaluate (see (\ref{12fintegral})-(\ref{CU}))
\bea
\label{U}
U_{\rm f} &=& \int d\tau_{1>2>3}  \, \Big\{ (\eta_2 \g_\mu \bar{\eta}_3) \varepsilon_{\nu \rho \s} \, \dot{x}_1^\nu \Big(  \Gamma^{\s\rho\mu} + \Gamma^{\s\mu\rho} \Big)
\\
&~& - (\eta_3 \g_\mu \bar{\eta}_1) \varepsilon_{\nu \rho \s} \, \dot{x}_2^\nu \Big(  \Gamma^{\s\rho\mu} + \Gamma^{\s\mu\rho} \Big)
+ (\eta_1 \g_\mu \bar{\eta}_2) \varepsilon_{\nu \rho \s} \, \dot{x}_3^\nu \Big(  \Gamma^{\s\rho\mu} + \Gamma^{\s\mu\rho} \Big) \Big\} \non 
\eea
The procedure required is quite complicated and far from being straightforward. We proceed step by step by collecting the most technical details in Appendices E and F, in order not to compromise the comprehension. 

As described in details in Appendix \ref{scalarintegral} (see (\ref{E4})-(\ref{E8})) each of these terms can be manipulated as
\begin{equation}\label{diminustwofactor}
(\eta_i \g_\mu \bar{\eta}_j) \varepsilon_{\nu \rho \s} \, \dot{x}_k^\nu \Big(  \Gamma^{\s\rho\mu} + \Gamma^{\s\mu\rho}   \Big) = - 4 (D-2) \sin \frac{\tau_{ij}}{2} \left(\sin ^2\frac{\tau_{ki}}{2} + \sin ^2\frac{\tau_{kj}}{2} \right)  I(x)
\end{equation}
where $I(x)$ is the scalar triangle integral
\beq
I(x)=\frac{4\pi^{3/2-\epsilon}\,\Gamma(2-2\epsilon)}{\Gamma^3(\tfrac{1}{2}-\epsilon)}
\int [d\alpha]
\frac{(\alpha_1\alpha_2\alpha_3)^{1/2-\epsilon}}
{\left( \alpha_1\alpha_2 x_{12}^2+\alpha_2\alpha_3 x_{23}^2+\alpha_1\alpha_3 x_{13}^2 \right)^{2-2\epsilon}}
\eeq
The overall factor $(D-2)= (1-2\e)$ in (\ref{diminustwofactor}) arises when contracting $\varepsilon$ tensors in DRED, as detailed in Appendix \ref{scalarintegral}. 
Its appearance is crucial in our computation. In fact, since it contains an evanescent term, it gives rise to non--trivial finite contributions when multiplied by divergent integrals.   

Now, using the the Mellin--Barnes representation (\ref{eq:MB}) for the $I(x)$ integral, introducing the convenient notation
\begin{equation}
{\cal S}(\alpha,\beta,\gamma)=
\left[\sin ^{2}\left(\frac{\tau _{12}}{2}\right)\right]^\alpha 
\left[\sin ^{2}\left(\frac{\tau _{23}}{2}\right)\right]^\beta
\left[\sin ^{2}\left(\frac{\tau _{13}}{2}\right)\right]^\gamma
\end{equation}
and observing that on the ordered contour of integration all the sines are always positive, we turn $U_{\rm f}$ to the following form
\begin{align}\label{ProblemToTackle}
& U_{\rm f} = -\frac{2^{4\epsilon}\, \pi^{\frac32-\epsilon}(1-2\epsilon)}{\Gamma^3 \left( \frac12 - \epsilon \right) \Gamma\left( \frac12 + \epsilon \right)}\, 
\int \frac{du\,dv}{(2\pi i)^2}\, \Gamma (-u) \Gamma (-v) \Gamma \left(-u\!+\!\epsilon\! -\!\frac{1}{2}\right) 
\\&
\phantom{dejo espacio} \Gamma \left(-\!v\!+\!\epsilon\! -\!\frac{1}{2}\right) \Gamma (u\!+\!v\!-\!2 \epsilon\! +\!2) \Gamma \left(u\!+\!v\!-\!\epsilon\! +\!\frac{3}{2}\right)
\; \times  
\nonumber\\ & 
\int _0^{2 \pi }\!\!\!\!\int _0^{\tau_1}\!\!\!\!\int _0^{\tau_2}d\tau_3 d\tau_2 d\tau_1 \, \left[ {\cal S}(u\!+\!1,v\!+\!\tfrac{1}{2},-\!u\!-\!v\!-\!2\!+\!2\epsilon) + {\cal S}(u\!+\!\tfrac{1}{2},v\!+\!1,-\!u\!-\!v\!-\!2\!+\!2\epsilon)  \right.
\nonumber\\ &
\phantom{dejo espacio} \phantom{dejo espacio} 
~ + \left. {\cal S}(u\!+\!1,v,-\!u\!-\!v\!-\!\tfrac{3}{2}\!+\!2\epsilon) +{\cal S}(u,v\!+\!1,-\!u\!-\!v\!-\!\tfrac{3}{2}\!+\!2\epsilon) \right.
\nonumber\\
& \left. 
\phantom{dejo espacio} \phantom{dejo espacio} 
+{\cal S}(u\!+\!\tfrac{1}{2},v,-\!u\!-\!v\!-\!1\!+\!2\epsilon) +{\cal S}(u,v\!+\!\tfrac{1}{2},-\!u\!-\!v\!-\!1\!+\!2\epsilon) \right] \non
\end{align}
Exploiting the possibility of performing change of variables in the  Mellin--Barnes integrations, it is not difficult to conclude that the total integrand is symmetric under exchanges of any couple of contour parameters $\tau_i$. For instance, the $\tau_1\leftrightarrow\tau_2$ exchange accompanied by the  shift $v\rightarrow -u-v+2\epsilon-2$ maps the integral into itself, as well as 
the exchange $\tau_2\leftrightarrow\tau_3$ together with the shift $u\rightarrow -u-v+2\epsilon-2$ and $\tau_1\leftrightarrow\tau_3$ with the relabelling $u\leftrightarrow v$. As a nice consequence, we can symmetrize the contour of integration as usual by replacing 
\begin{equation}
\int _0^{2 \pi } \!\!\!\!d\tau_1  \int _0^{\tau_1} \!\!\!\!d\tau_2  \int _0^{\tau_2} \!\!\!\!d\tau_3  
\rightarrow \frac{1}{3!}
\int _0^{2 \pi } \!\!\!\!d\tau_1 \int _0^{2\pi} \!\!\!\!d\tau_2 \int _0^{2\pi} \!\!\!\!d\tau_3  
\end{equation}
and we are left with the following expression to be evaluated
\begin{align}
\label{U2}
& U_{\rm f} = -\frac{2^{4\epsilon}\, \pi^{\frac32-\epsilon} (1-2\epsilon)}{3!\,\Gamma^3 \left( \frac12 - \epsilon \right) \Gamma\left( \frac12 + \epsilon \right)}\, \int \frac{du\,dv}{(2\pi i)^2}\, \Gamma (-u) \Gamma (-v) \Gamma \left(-\!u\!+\!\epsilon\! -\!\frac{1}{2}\right) \Gamma \left(-\!v\!+\!\epsilon\! -\!\frac{1}{2}\right)
\nonumber\\&
\Gamma (u\!+\!v\!-\!2 \epsilon\! +\!2) \Gamma \left(u\!+\!v\!-\!\epsilon\! +\!\frac{3}{2}\right)
\, \left[ \mathcal{J}(u\!+\!1,v\!+\!\tfrac{1}{2},-\!u\!-\!v\!-\!2\!+\!2\epsilon) +
\right. \nonumber\\ &
\phantom{dejo espacio}
+\mathcal{J}(u\!+\!\tfrac{1}{2},v\!+\!1,-\!u\!-\!v\!-\!2\!+\!2\epsilon) 
+\mathcal{J}(u\!+\!1,v,-\!u\!-\!v\!-\!\tfrac{3}{2}\!+\!2\epsilon)  +
\nonumber\\ &
\phantom{dejo espacio}
+\mathcal{J}(u,v\!+\!1,-\!u\!-\!v\!-\!\tfrac{3}{2}\!+\!2\epsilon)
+\mathcal{J}(u\!+\!\tfrac{1}{2},v,-\!u\!-\!v\!-\!1\!+\!2\epsilon) + \nonumber\\
& \left.
\phantom{dejo espacio}
+\mathcal{J}(u,v\!+\!\tfrac{1}{2},-\!u\!-\!v\!-\!1\!+\!2\epsilon) \right]
\end{align}
where $\mathcal{J}$ is the symmetrized integral
\begin{equation}
\label{J}
\mathcal{J}(\alpha,\beta,\gamma)=
\int _0^{2 \pi }\!\!\!\!d\tau_1\int _0^{2\pi}\!\!\!\!d\tau_2\int _0^{2\pi}\!\!\!\!d\tau_3
\left[\sin ^{2}\left(\frac{\tau _{12}}{2}\right)\right]^\alpha
\left[\sin ^{2}\left(\frac{\tau _{23}}{2}\right)\right]^\beta
\left[\sin ^{2}\left(\frac{\tau _{13}}{2}\right)\right]^\gamma
\end{equation}
The details of the calculation can be found in Appendix \ref{app:uncontracted}. The final result for $\mathcal{J}(\alpha,\beta,\gamma)$ reads
\begin{equation}
\mathcal{J}(\alpha,\beta,\gamma)=
8\pi^{3/2} \,
\frac{\Gamma(\tfrac{1}{2}+\alpha)\Gamma(\tfrac{1}{2}+\beta)\Gamma(\tfrac{1}{2}+\gamma)\Gamma(1+\alpha+\beta+\gamma)}
{\Gamma(1+\alpha+\gamma)\Gamma(1+\beta+\gamma)\Gamma(1+\alpha+\beta)}
\end{equation}
Plugging this into (\ref{U2}) the uncontracted integrals can be written as 
\begin{align}
& U_{\rm f} = -\frac{2^{4\epsilon+3}\, \pi^{3-\epsilon} (1-2\epsilon) \Gamma \left(\frac{1}{2}+ 2 \epsilon \right)}
{\,\Gamma^3 \left( \frac12 - \epsilon \right) \Gamma\left( \frac12 + \epsilon \right)}
\ G(1,\tfrac{1}{2})
\end{align}
where $G(1,\tfrac{1}{2})$ is a Mellin--Barnes integral, which can be evaluated by expanding the integrand in powers of $\epsilon$, so obtaining
(see Appendix \ref{app:uncontracted} for details) 
\begin{equation}
G(1,\tfrac{1}{2})= \frac{\pi^{3/2}}{2}\, \left( \frac{1}{\epsilon} - 4 + 3 \gamma_E + 6 \log 2 \right)
\end{equation}
The $\frac{1}{\epsilon}$ pole signals the presence of a short distance divergence at this intermediate stage. As we are going to show, this divergence gets cancelled by an analogous contribution coming from the contracted integrals $C_{\rm f}$. However, it plays an important role in determining the final result since, being multiplied by the factor $(1-2\e)$ arising from the application of DRED rules, it contributes non--trivially with finite terms which survive the physical limit $\e \to 0$. 

The final result for the uncontracted part of diagram 2(f) expanded up to finite terms is
\begin{equation}\label{eq:uncontracted}
\boxed{ U_{\rm f} = 4\pi^3
\left(-\frac{1}{\epsilon}+6+\gamma_E-2\log 2 +\log \pi \right) } + {\cal O}(\epsilon)
\end{equation}

\subsection{The contracted integrals  $C_{\rm f}$}

We now concentrate on the evaluation of $C_{\rm f}$, coming from the sum of the last three terms in (\ref{12fintegral}) plus their permutations. 

Tensor integrals $\G^{\mu\nu\rho}$ in eq. (\ref{gammaintegrals}) are immediately performed when a pair of indices are contracted. In fact, in that case they are proportional to the operator $\pa^\mu \pa_\mu$ acting on a scalar propagator in the integrand, so we can make use of the Green equation
\begin{equation}
\Box\, \frac{1}{(x_{ij}^2)^{1/2-\e}} = -\frac{4\pi^{3/2-\e}}{\Gamma\left(\frac12-\e\right)}\, \d (x_{ij})
\end{equation}
For the three structures appearing in (\ref{12fintegral}) we find
\begin{align}
&\Gamma^{\nu\mu}_{\phantom{\nu\mu}\mu} = -\frac{2\pi^{3/2-\e}}{\Gamma\left(\frac12-\e\right)}\, \partial_1^{\nu} \,\left[ \frac{1}{(x_{12}^2)^{\tfrac12-\e}(x_{13}^2)^{\tfrac12-\e}} - \frac{1}{(x_{12}^2)^{\tfrac12-\e}(x_{23}^2)^{\tfrac12-\e}} - \frac{1}{(x_{13}^2)^{\tfrac12-\e}(x_{23}^2)^{\tfrac12-\e}} \right]
\nonumber \\&
\Gamma^{\mu}_{\phantom{\mu}\nu\mu} = -\frac{2\pi^{3/2-\e}}{\Gamma\left(\frac12-\e\right)}\, \partial_2^{\nu} \,\left[ \frac{1}{(x_{12}^2)^{\tfrac12-\e}(x_{23}^2)^{\tfrac12-\e}} - \frac{1}{(x_{13}^2)^{\tfrac12-\e}(x_{23}^2)^{\tfrac12-\e}} - \frac{1}{(x_{12}^2)^{\tfrac12-\e}(x_{13}^2)^{\tfrac12-\e}} \right]
\nonumber \\&
\Gamma^{\mu}_{\phantom{\mu}\mu\nu} = -\frac{2\pi^{3/2-\e}}{\Gamma\left(\frac12-\e\right)}\, \partial_3^{\nu} \,\left[ \frac{1}{(x_{13}^2)^{\tfrac12-\e}(x_{23}^2)^{\tfrac12-\e}} - \frac{1}{(x_{12}^2)^{\tfrac12-\e}(x_{13}^2)^{\tfrac12-\e}} - \frac{1}{(x_{12}^2)^{\tfrac12-\e}(x_{23}^2)^{\tfrac12-\e}} \right]
\end{align}

Parametrizing the coordinates on the unit circle, exploiting the identities (\ref{eq:etaetabar}, \ref{eq:etagamma0}) and summing over all contributions, after a quite lengthy calculation we obtain
\begin{align}
\label{eq:intfcontr}
C_{\rm f} & = 
\frac{(1-2\e) \pi^{\frac32 - \e}}{2^{1-4\e} \G(\frac12 -\e)} \, \int_0^{2\pi}   d \tau_1 \int_0^{\tau_1} d\tau_2 \int_0^{\tau_2} d\tau_3 \; \times
\\ 
& \left\{ -\frac12 \frac{\sin \left(\tau _{13}\right)}{\left( \sin\left(\frac{\tau _{12}}{2}\right) \sin\left(\frac{\tau _{23}}{2}\right) \right)^{2-2\epsilon} }
+ \frac12 \frac{\sin \left(\tau _{12}\right)}{\left( \sin\left(\frac{\tau _{13}}{2}\right) \sin\left(\frac{\tau _{23}}{2}\right) \right)^{2-2\epsilon} }
+\frac12 \frac{\sin \left(\tau _{23}\right)}{\left( \sin\left(\frac{\tau _{12}}{2}\right) \sin\left(\frac{\tau _{13}}{2}\right) \right)^{2-2\epsilon} } \right.
\non \\
\non \\
& \left. - \frac{2\sin \left(\frac{\tau _{23}}{2}\right)}{\left( \sin\left(\frac{\tau _{12}}{2}\right) \sin\left(\frac{\tau _{13}}{2}\right) \right)^{1-2\epsilon}}
- \frac{2\sin \left(\frac{\tau _{13}}{2}\right)}{\left( \sin\left(\frac{\tau _{12}}{2}\right) \sin\left(\frac{\tau _{23}}{2}\right) \right)^{1-2\epsilon}}
- \frac{2\sin \left(\frac{\tau _{12}}{2}\right)}{\left( \sin\left(\frac{\tau _{13}}{2}\right) \sin\left(\frac{\tau _{23}}{2}\right) \right)^{1-2\epsilon}} \right.
\non \\
\non \\
& \left.  + \frac{\cos \left(\frac{\tau _{23}}{2}\right)}{  \sin ^{2-2\epsilon}\left(\frac{\tau _{12}}{2}\right) \sin ^{1-2 \epsilon}\left(\frac{\tau _{13}}{2}\right)}
+ \frac{\cos \left(\frac{\tau _{23}}{2}\right)}{ \sin ^{1-2 \epsilon}\left(\frac{\tau _{12}}{2}\right) \sin ^{2-2\epsilon}\left(\frac{\tau _{13}}{2}\right)}
- \frac{\cos \left(\frac{\tau _{13}}{2}\right)}{\sin ^{1-2 \epsilon}\left(\frac{\tau _{12}}{2}\right) \sin ^{2-2\epsilon}\left(\frac{\tau _{23}}{2}\right)} \right.
\non \\
\non \\
& \left. + \frac{\cos \left(\frac{\tau _{12}}{2}\right)}{\sin ^{2-2\epsilon}\left(\frac{\tau _{13}}{2}\right) \sin ^{1-2 \epsilon}\left(\frac{\tau _{23}}{2}\right)}
+ \frac{\cos \left(\frac{\tau _{12}}{2}\right)}{\sin ^{1-2 \epsilon}\left(\frac{\tau _{13}}{2}\right) \sin ^{2-2\epsilon}\left(\frac{\tau _{23}}{2}\right)} 
- \frac{\cos \left(\frac{\tau _{13}}{2}\right)}{ \sin ^{2-2\epsilon}\left(\frac{\tau _{12}}{2}\right) \sin ^{1-2 \epsilon}\left(\frac{\tau _{23}}{2}\right)}
 \right\}
\non
\end{align}
These integrals can be computed by expanding the trigonometric functions in power series and following the method described in Appendix \ref{sec:method}. 
It turns out that, once expressed as series, six of these expressions reduce to the others, so that it is sufficient to compute six independent integrals out of twelve. 
Nonetheless, the calculation is quite involved; we detail it in Appendix \ref{contracted}, while here we simply quote the final result 
\begin{align}
C_{\rm f} &= \frac{2 \pi^{\frac32-\e} (1-2\e)}{\G \left( \frac12 -\e \right)} \; \times  
\non \\
&  \Big\{ 8 \frac{ \pi ^2 (\epsilon -1) \epsilon  (2 \epsilon -1) \Gamma^2 (-1+2 \epsilon)}{\Gamma^4 (1+\epsilon)} -\frac{  2^{1+4 \epsilon} \, \pi ^{3/2} \, \Gamma \left(\frac12 + 2 \epsilon \right)}{\epsilon ^2 \Gamma (2 \epsilon )} 
\non \\
& +8\frac{ \pi  \sin (\pi  \epsilon ) \Gamma (2-\epsilon ) \Gamma (2 \epsilon ) \Gamma (-1+2 \epsilon)}{\Gamma^3 (1+\epsilon)} + \frac{2^{1+4\e}\, \pi ^2 \, \left(\sin ^2(2 \pi  \epsilon )-2 \cos (2 \pi  \epsilon )\right)}{\epsilon\, \sin^2(2 \pi  \epsilon ) }
\non \\
& + \frac{4 \pi ^2 }{2-\epsilon } \; 
{}_3F_2\left[
             \begin{array}{c}
             2-2 \epsilon ,2-2 \epsilon ,2-\epsilon \\
             1,3-\epsilon \\
             \end{array};1
             \right]
+  \frac{4 \pi ^2 }{\e}   \, 
{}_3F_2\left[
             \begin{array}{c}
             2-2 \epsilon ,2-2 \epsilon ,-\epsilon \\
             1,1-\epsilon \\
             \end{array};1
             \right]
\nonumber
\end{align}
\begin{align}
\label{intfcontr}
&+ 8 \frac{ \cos (2 \pi  \epsilon ) \Gamma \left(\tfrac52-2 \epsilon\right) \Gamma (-1+2 \epsilon) }{\sqrt{\pi } \epsilon^2 (4 \epsilon -3)} \, 
{}_4F_3\left[
             \begin{array}{c}
             1-2 \epsilon ,\frac{3}{2}-2 \epsilon ,-\epsilon ,-\epsilon \\
             \frac{1}{2},1-\epsilon ,1-\epsilon \\
             \end{array};1
             \right]
\nonumber\\&
+8 \frac{ \cos (2 \pi  \epsilon ) \Gamma \left(\tfrac52-2 \epsilon\right) \Gamma (-1+2 \epsilon)}{\sqrt{\pi } (\epsilon -1)^2 (4 \epsilon -3)} \, 
{}_4F_3\left[
             \begin{array}{c}
             1-2 \epsilon ,\frac{3}{2}-2 \epsilon ,1-\epsilon ,1-\epsilon \\
             \frac{1}{2},2-\epsilon ,2-\epsilon \\
             \end{array};1
             \right]
 \non \\&
-32 \frac{\sqrt{\pi } \Gamma (2 \epsilon )}{(2 \epsilon -1)^2 \, \Gamma \left( -\frac{1}{2} + 2 \epsilon \right)} \, 
{}_4F_3\left[
             \begin{array}{c}
             \frac{3}{2}-2 \epsilon ,2-2 \epsilon ,\frac{1}{2}-\epsilon ,\frac{1}{2}-\epsilon \\
             \frac{3}{2},\frac{3}{2}-\epsilon ,\frac{3}{2}-\epsilon \\
             \end{array};1
             \right]
\nonumber\\&
-32 \frac{ \sqrt{\pi } \Gamma (2 \epsilon )}{(2 \epsilon -3)^2 \, \Gamma \left(-\frac12 + 2 \epsilon \right)} \,
{}_4F_3\left[
             \begin{array}{c}
             \frac{3}{2}-2 \epsilon ,2-2 \epsilon ,\frac{3}{2}-\epsilon ,\frac{3}{2}-\epsilon \\
             \frac{3}{2},\frac{5}{2}-\epsilon ,\frac{5}{2}-\epsilon \\
             \end{array};1
             \right]
 \Big\}
\nonumber\\
\end{align}
given in terms of generalized hypergeometric functions, ${}_pF_q$. The defining series for the ${}_3F_2$ and ${}_4F_3$ functions appearing here converge for the half--plane $\Re(\epsilon)>1/2$ and $\Re(\epsilon)>0$ respectively. Consequently, expression (\ref{intfcontr}) is not well--defined close to $\e = 0$. As anticipated, the general strategy for continuing it to that region and performing the $\e$--expansion will be given in the next Section. \\

\section{Analytic continuations}\label{sec:expansions}

We now address the problem of expanding the expressions (\ref{eq:I2f}, \ref{intfcontr}) in powers of $\epsilon$ around $\epsilon = 0$, which is the physical limit we are interested in.
Looking at the structure of the corresponding integrals it is easy to realize that many hypergeometric functions appearing there are well--defined only in regions of the complex plane that do not include the origin.
In the spirit of dimensional regularization, we analytically continue these functions to include at least a neighbourhood of $\epsilon =0$, and then expand in powers of $\e$.

\subsection{General procedure}

To carry out such a project we first observe that our results involve two prototypes of hypergeometric functions  
\begin{equation}\label{eq:omega}
\Omega_1(\epsilon)={}_3F_2\left[
                            \begin{array}{c}
                              a,b,d \\
                              c,d\!\!+\!\!1 \\
                            \end{array};1
                          \right]
\quad\mbox{and}\quad
\Omega_2(\epsilon)={}_4F_3\left[
                            \begin{array}{c}
                              a, b, d, d \\
                              c,d\!\!+\!\!1,d\!\!+\!\!1 \\
                            \end{array};1
                          \right]
\end{equation}
where $a$, $b$, $c$ and $d$ are linear functions of $\epsilon$. Both hypergeometric series converge in the half--plane $\Re(1+c-a-b)>0$.

As already outlined, the particular values of $a,b,c,d$ involved in the results (\ref{eq:I2f}, \ref{intfcontr}) imply convergence of the corresponding series away from $\epsilon=0$. Therefore, taking the straight series definition of the hypergeometrics, and making the expansion for small $\epsilon$ is ill--defined and analytic continuation around $\epsilon=0$ is then required.

By applying few algebraic transformations on the original hypergeometric functions we manage to express them in terms of extended hypergeometric functions and their derivatives. These generalized functions match the original ones in the domain of the complex $\epsilon$--plane where they converge, but they are well-defined in a greater region, including a neighbourhood of $\epsilon=0$.

We begin by considering $\Omega_1(\epsilon)$. Its series definition reads
\begin{equation}
\Omega_1(\epsilon)= d \, \frac{ \Gamma(c)}{\Gamma(a)\Gamma(b)}
\sum\limits_{n=0}^{\infty}\frac{\Gamma(a+n)\Gamma(b+n)}{n!\,\Gamma(c+n)}
\frac{1}{(d+n)}
\end{equation}
We can Schwinger--parametrize the rational piece inside the series
\begin{equation}
\frac{1}{(d+n)}=
\int\limits_{0}^{\infty}dt\, e^{-t(n+d)}
\end{equation}
to obtain
\begin{equation}
\Omega_1(\epsilon)=
d \int\limits_{0}^{\infty}dt\, e^{-t d}\,
                          {}_2F_1\left[
                            \begin{array}{c}
                              a,b \\
                              c \\
                            \end{array};e^{-t}
                          \right]
\end{equation}
We now use Euler fractional linear transformation of the Gau{\ss} hypergeometric ${}_2F_1$ function\footnote{See Abramowitz \& Stegun, page 559.}
\begin{equation}
                           {}_2F_1\left[
                            \begin{array}{c}
                              a,b \\
                              c \\
                            \end{array}; z
                          \right]
=(1-z)^{c-a-b}\,\,
                         {}_2F_1\left[
                            \begin{array}{c}
                              c\!\!-\!\!a,c\!\!-\!\!b \\
                              c \\
                            \end{array}; z
                          \right]
\end{equation}
to obtain
\begin{equation}
\Omega_1(\epsilon)=
d \, \frac{\Gamma(c)}{\Gamma(c-a)\Gamma(c-b)}
\sum\limits_{n=0}^{\infty}\frac{\Gamma(c-a+n)\Gamma(c-b+n)}{n!\,\Gamma(c+n)}
\int\limits_{0}^{\infty}dt\,e^{-t(n+d)}(1-e^{-t})^{c-a-b}.
\end{equation}
Integrating in $t$ we have the new series definition for $\Omega_1(\epsilon)$
\begin{equation}
\label{omega1}
\Omega_1(\epsilon)=
d \, \frac{\Gamma(1-a-b+c)\Gamma(c)}{\Gamma(c-a)\Gamma(c-b)}
\sum\limits_{n=0}^{\infty}\frac{\Gamma(c-a+n)\Gamma(c-b+n)\Gamma(d+n)}
{n!\,\Gamma(c+n)\Gamma(1-a-b+c+d+n)}
\end{equation}

Similarly, for $\Omega_2(\epsilon)$ we write the defining series
\begin{equation}
\Omega_2(\epsilon)=  d^2 \, \frac{ \Gamma(c)}{\Gamma(a)\Gamma(b)}
\sum\limits_{n=0}^{\infty}\frac{\Gamma(a+n)\Gamma(b+n)}{n!\,\Gamma(c+n)}
\frac{1}{(d+n)^2}
\end{equation}
We can again Schwinger--parametrize the rational piece inside the series
\begin{equation}
\frac{1}{(d+n)^2}=
\int\limits_{0}^{\infty}dt\,t\, e^{-t(n+d)}
\end{equation}
to obtain
\begin{equation}
\Omega_2(\epsilon)=
d^2 \int\limits_{0}^{\infty}dt\,t\, e^{-t d}\,
                          {}_2F_1\left[
                            \begin{array}{c}
                              a,b \\
                              c \\
                            \end{array};e^{-t}
                          \right]
\end{equation}
Once again, Euler fractional linear transformation of the Gau{\ss} hypergeometric ${}_2F_1$ function allows us to write
\begin{equation}
\label{omega2}
\Omega_2(\epsilon)=
d^2 \, \frac{\Gamma(c)}{\Gamma(c-a)\Gamma(c-b)}
\sum\limits_{n=0}^{\infty}\frac{\Gamma(c-a+n)\Gamma(c-b+n)}{n!\,\Gamma(c+n)}
\int\limits_{0}^{\infty}dt\,t\,e^{-t(n+d)}(1-e^{-t})^{c-a-b}
\end{equation}
Integrating in $t$ we have the new series definition for $\Omega_2(\epsilon)$
\begin{align}
\label{eq:omega2}
&\Omega_2(\epsilon)=
d^2 \, \frac{\Gamma(1-a-b+c)\Gamma(c)}{\Gamma(c-a)\Gamma(c-b)}\times \\
\sum\limits_{n=0}^{\infty}&\frac{\Gamma(c-a+n)\Gamma(c-b+n)\Gamma(d+n)}
{n!\,\Gamma(c+n)\Gamma(1-a-b+c+d+n)}
\left(\psi^{(0)}(1-\!a\!-\!b\!+\!c\!+\!d\!+\!n)-\psi^{(0)}(d\!+\!n)\right) \nonumber
\end{align}
where $\psi^{(0)}$ is the digamma function (see eq. (\ref{gammader})).

Studying the convergence of the new series (\ref{omega1}, \ref{eq:omega2}) shows that they are well--defined for any value of $a$, $b$, $c$ and $d$; therefore they admit a Laurent expansion centered in $\epsilon=0$ which is what we intended to obtain.

When applying these techniques to our case, it turns out that although most of the series needed to complete the computation can be found in literature, few of them cannot and require a direct evaluation. We list them in Appendix \ref{app:series}.

The technique just explained for analytic continuation can be applied to more general hypergeometric functions \footnote{See for instance Bailey, Generalized hypergeometric series, pag. 98.}, but this is beyond the scope of the present analysis.

\subsection{Specific results for diagrams 2(e), 2(f)}

Equipped with these tools, we now perform analytic continuation of the hypergeometric functions appearing in the results (\ref{eq:I2f}, \ref{intfcontr}).

\paragraph{Diagram 2(e): Result (\ref{eq:I2f}).}

While the $\Gamma$ functions appearing in (\ref{eq:I2f}) can be easily expanded in powers of $\epsilon$, the hypergeometric function has to be analytically continued using the above procedure.
We rewrite the first line in (\ref{eq:I2f}) as 
\beq
 - 8 \pi^{1+2\e} \frac{\G^2(\frac32 - \e)}{\G^2(2-2\e)} \rho_1(\e)
\eeq 
where we have defined $\rho_1$ to be the hypergeometric series 
\begin{equation}
\rho_1(\epsilon) \equiv  
\frac{\Gamma^2(2-2\epsilon)}{(1-\epsilon)^2}\,
{}_4F_3\left[
             \begin{array}{c}
             2\!-\!2\epsilon,2\!-\!2\epsilon,1-\!\epsilon,1-\!\epsilon \\
             1,2\!-\!\epsilon,2\!-\!\epsilon \\
             \end{array};1
             \right] = \sum\limits_{n=0}^{\infty}
\frac{\Gamma^2(2-2\epsilon+n)}{(n!)^2 (1-\epsilon+n)^2}
\end{equation}
The defining series of this function is convergent for the half--plane $\Re(\epsilon)>1/2$. It belongs to the class $\Omega_2$ defined in (\ref{eq:omega}), hence we can apply formula (\ref{eq:omega2}) to perform analytic continuation. We then write
\begin{align}
\rho_1(\epsilon)&=
\frac{\Gamma^2 (2\!-\!2 \epsilon )\Gamma (-2 \!+\!4 \epsilon)}
{\Gamma^2 (-1 \!+\! 2 \epsilon)}
\sum\limits_{n=0}^{\infty}
\frac{\Gamma (n\!-\!\epsilon +1) \Gamma^2 (n\!+\!2 \epsilon\!-\!1) \left(\psi^{(0)}(n\!+\!3 \epsilon\!-\!1)-\psi^{(0)}(n\!-\!\epsilon\!+\!1)\right)}{(n!)^2 \, \Gamma (n\!+\!3\epsilon\!-\!1)}
\end{align}
Expanding around $\epsilon=0$ we need take special care of the first two terms of the series, namely the ones corresponding to $n=0$ and $n=1$, which develop simple poles in $\epsilon$. These singularities are however cancelled by $\e$ contributions from the overall factor in front. Taking into account that the remaining series contributes only to 
$\mathcal{O}(\epsilon)$, we obtain
\begin{equation}
\label{rho1}
\rho_1(\epsilon) = -\frac{1}{2}-(1+2\gamma ) \epsilon +\mathcal{O}\left(\epsilon^2\right)
\end{equation}
and the contribution from diagram 2(e) finally reads
\begin{equation}
\label{eq:Ieexpanded}
\boxed{\boxed{{\rm (e)} = \frac{3}{2} \pi^2 \, \frac{MN}{k^2} }} + \mathcal{O}(\epsilon)
\end{equation}

\paragraph{Diagram 2(f): Result  (\ref{intfcontr}).}

There are six hypergeometric functions in the contracted part of diagram 2(f), eq.  (\ref{intfcontr}), that require analytic continuation. The two ${}_3F_2$ ones have defining series which converge for $\Re(\epsilon)>1/2$ while the four ${}_4F_3$ ones converge for $\Re(\epsilon)>0$. We analyze them separately, following the order in which they appear in the result.

\vskip 5pt
\noindent
$\bullet$ First, we consider
\begin{equation}
\rho_2(\epsilon) \equiv  \frac{4\pi^2 }{2-\epsilon}\;
{}_3F_2\left[
             \begin{array}{c}
             2\!-\!2\epsilon,2\!-\!2\epsilon,2\!-\!\epsilon \\
             1,3\!-\!\epsilon \\
             \end{array};1
             \right]
\end{equation}
The series defining this hypergeometric function converges for $\Re(\epsilon)>\tfrac{1}{2}$. Applying (\ref{omega1}) we obtain
\begin{equation}
\rho_2(\epsilon)=
\frac{4 \pi ^2 \Gamma(-2+4 \epsilon)}{\Gamma^2(-1+2 \epsilon)}
\sum\limits_{n=0}^{\infty}
\frac{\Gamma(2+n-\epsilon) \Gamma^2(-1+n+2 \epsilon) }{(n!)^2 \Gamma(n+3 \epsilon)}.
\end{equation}
Isolating the first two divergent terms in the sum as done in the previous case and $\e$--expanding the argument of the remaining series we can write 
\begin{align}
&\rho_2(\epsilon)=\frac{4 \pi ^2 \Gamma(2-\epsilon ) \Gamma(-2+4 \epsilon )}{\Gamma(3 \epsilon )}+
\frac{2 \pi ^2 (-1+2 \epsilon ) \Gamma(3-\epsilon) \Gamma(-1+4 \epsilon)}{\Gamma(1+3 \epsilon )}\nonumber\\
&+2\pi^2\epsilon\sum\limits_{n=2}^{\infty}
\frac{n + 1}{n (n-1)^2 }+\mathcal{O}(\epsilon^2)
\end{align}
Finally, summing the series, we obtain
\begin{equation}
\rho_2(\epsilon)=\frac{\pi ^2}{\epsilon }+2 \pi ^2+\left(11 \pi ^2+\frac{4 \pi ^4}{3}\right) \epsilon +\mathcal{O}(\epsilon^2)
\end{equation}

\vskip 8pt
\noindent
$\bullet$ Next we analyze
\begin{equation}
\rho_3(\epsilon) \equiv  \frac{4\pi^2}{\epsilon}\
{}_3F_2\left[
             \begin{array}{c}
             2\!-\!2\epsilon,2\!-\!2\epsilon,-\!\epsilon \\
             1,1\!-\!\epsilon \\
             \end{array};1
             \right]
\end{equation}
Applying (\ref{omega1}) yields
\begin{equation}
\rho_3(\epsilon)=
-\frac{4 \pi^2 \Gamma(-2+4 \epsilon)}{\Gamma^2(-1+2 \epsilon)}
\sum\limits_{n=0}^{\infty}
\frac{ \Gamma(n-\epsilon) \Gamma^2(-1+n+2 \epsilon)}{(n!)^2 \Gamma(-2+n+3 \epsilon)}
\end{equation}
and its expansion reads
\begin{equation}
\rho_3(\epsilon)=\frac{3 \pi ^2}{\epsilon }+6 \pi ^2+\left(21 \pi ^2+\frac{4 \pi ^4}{3}\right) \epsilon +\mathcal{O}(\epsilon^2)
\end{equation}

We now turn to the four contributions proportional to the $_4 F_3$ hypergeometric series, which have more complicated expansions.  

After applying prescription (\ref{eq:omega2}), we need sum series whose generic term is a function 
of $\e$. 
Similarly to the previous situations, in general it occurs that the first few terms of the series develop $\e$--pole singularities, so we analyze them separately. In the remaining series we expand the summand in powers of $\e$. Given that some of their overall coefficients have poles in $\epsilon$, we have to expand up to higher orders. This converts the original series into the linear combination of a finite number of series that contain higher polygamma functions $\psi^{(n)}$ defined in (\ref{gammader}). Closer we go to $\mathcal{O}(\epsilon^0)$, more complicated these series become. Therefore, while the evaluation of the first few ones is relatively easy and can be done by {\em Mathematica}, the rest has required explicit evaluation. The results are listed in Appendix G.

\vskip 8pt
\noindent
$\bullet$ We consider the first term proportional to $_4 F_3$ appearing in (\ref{intfcontr})
\begin{equation}
\rho_4(\epsilon)=
8 \frac{ \cos (2 \pi  \epsilon ) \Gamma\left(\tfrac52-2 \epsilon\right) \Gamma (-1+2 \epsilon) }{\sqrt{\pi } \epsilon^2 (4 \epsilon -3)}\,
{}_4F_3\left[
             \begin{array}{c}
             1\!-\!2\epsilon, 3/2\!-\!2\epsilon,-\!\epsilon,-\!\epsilon \\
             1/2,1-\!\epsilon,1-\!\epsilon \\
             \end{array};1
             \right]
\end{equation}
Its analytic continuation gives
\begin{align}
&\rho_4(\epsilon)=\frac{\pi ^{3/2} 4^{2 \epsilon} \csc (2 \pi  \epsilon )}{\Gamma (1-2 \epsilon ) \Gamma \left(-\frac{1}{2} + 2 \epsilon  \right)}
\times\nonumber\\
&\sum\limits_{n=0}^{\infty}
\frac{\Gamma (n\!-\!\epsilon ) \Gamma (n\!+\!2 \epsilon\! -\!1) \Gamma \left(n\!+\!2 \epsilon\! -\!\frac{1}{2}\right) (\psi ^{(0)}(n\!+\!3 \epsilon\! -\!1)-\psi ^{(0)}(n\!-\!\epsilon ))}{n! \, \Gamma \left(n\!+\!\frac{1}{2}\right) \Gamma (n\!+\!3 \epsilon \!-\!1)}
\end{align}
and its expansion around $\epsilon=0$ reads
\begin{align}
\rho_4(\epsilon)=&
\frac{1}{\epsilon ^3}+\frac{4\log 2-\frac{3}{2}}{\epsilon ^2}+\frac{-\pi ^2-6 + 16 \log 2 (\log 2-1)}{2 \epsilon }
\nonumber\\
&+\left(7 \zeta (3)+\frac{2}{3} \left(-9+ 4 \log^2 2 (4\log 2-9)-\pi ^2 (3\log 2-1)-24 \log 2 \right)\right)
\end{align}

\vskip 8pt
\noindent
$\bullet$ Then we consider the second term in (\ref{intfcontr}) proportional to $_4 F_3$ 
\begin{equation}
\rho_5(\epsilon)=
8 \frac{ \cos (2 \pi  \epsilon ) \Gamma \left(\tfrac52-2 \epsilon\right) \Gamma (-1 + 2 \epsilon) }{\sqrt{\pi } (\epsilon-1)^2 (4 \epsilon -3)}\,
{}_4F_3\left[
             \begin{array}{c}
             1\!-\!2\epsilon, 3/2\!-\!2\epsilon,1\!-\!\epsilon,1\!-\!\epsilon \\
             1/2,2\!-\!\epsilon,2\!-\!\epsilon \\
             \end{array};1
             \right]
\end{equation}
Its analytic continuation is
\begin{align}
&\rho_5(\epsilon)=
\frac{\pi ^{3/2} 4^{2 \epsilon} (1-2 \epsilon ) \epsilon ^2 \csc (2 \pi  \epsilon )}{(\epsilon -1)^2 \Gamma (2-2 \epsilon ) \Gamma \left(-\frac{1}{2} + 2 \epsilon  \right)}
\times\nonumber\\
&\times\sum\limits_{n=0}^{\infty}
\frac{\Gamma (n\!-\!\epsilon ) \Gamma (n\!+\!2 \epsilon\! -\!1) \Gamma \left(n\!+\!2 \epsilon\! -\!\frac{1}{2}\right) (\psi ^{(0)}(n\!+\!3 \epsilon\! -\!1)-\psi ^{(0)}(n\!-\!\epsilon ))}{n! \, \Gamma \left(n\!+\!\frac{1}{2}\right) \Gamma (n\!+\!3 \epsilon\! -\!1)}
\end{align}
and, consequently, its $\epsilon$--expansion reads
\begin{equation}
\rho_5(\epsilon)=
\frac{1}{2 \epsilon ^2} + \frac{1}{\epsilon } \left( 1 - \frac{\pi ^2}{6} \right)
+\left(7 \zeta (3) + 2-8 \log^2 2-\frac{2}{3} \pi^2 \log 2 \right)
\end{equation}

\vskip 8pt
\noindent
$\bullet$ The third term in (\ref{intfcontr}) proportional to $_4 F_3$ is 
\begin{equation}
\rho_6(\epsilon)=-32 \frac{\sqrt{\pi } \Gamma (2 \epsilon )}{(2 \epsilon -1)^2 \, \Gamma \left(-\frac{1}{2} + 2 \epsilon  \right)} \, 
{}_4F_3\left[
             \begin{array}{c}
             3/2\!-\!2\epsilon,2\!-\!2\epsilon,1/2-\!\epsilon,1/2-\!\epsilon \\
             3/2,3/2-\!\epsilon,3/2-\!\epsilon \\
             \end{array};1
             \right]
\end{equation}
Its analytical continuation reads
\begin{align}
&\rho_6(\epsilon)=
-\frac{\pi ^{3/2} 4^{2 \epsilon} \csc (2 \pi  \epsilon )}{\Gamma (1-2 \epsilon ) \Gamma \left(-\frac{1}{2} + 2 \epsilon \right)}
\times
\nonumber\\
&\times\sum\limits_{n=0}^{\infty}
\frac{\Gamma \left(n\!-\!\epsilon\!+\!\frac{1}{2}\right) \Gamma \left(n\!+\!2 \epsilon\! -\!\frac{1}{2}\right) \Gamma (n\!+\!2 \epsilon ) \left(\psi ^{(0)}\left(n\!+\!3 \epsilon\! -\!\frac{1}{2}\right)-\psi ^{(0)}\left(n\!-\!\epsilon\! +\!\frac{1}{2}\right)\right)}{n! \, \Gamma \left(n\!+\!\frac{3}{2}\right) \Gamma \left(n\!+\!3 \epsilon\! -\!\frac{1}{2}\right)}
\end{align}
allowing to obtain the expansion
\begin{align}
\rho_6(\epsilon)=&
\frac{1}{2 \epsilon^2}+\frac{2+\pi ^2+8\log 2}{2\epsilon }
\nonumber\\
&+\left(2 -7 \zeta (3)+16 \log^2 2+ 8\log 2+\frac{2}{3} \pi^2 (3\log 2-2)\right)+\mathcal{O}\left(\epsilon\right)
\end{align}

\vskip 8pt
\noindent
$\bullet$ Finally, the last term in (\ref{intfcontr}) is 
\begin{equation}
\rho_7(\epsilon)=
-32 \frac{\sqrt{\pi } \Gamma (2 \epsilon )}{(2 \epsilon -3)^2 \, \Gamma \left(-\frac{1}{2} + 2 \epsilon \right)} \, 
{}_4F_3\left[
             \begin{array}{c}
             3/2\!-\!2\epsilon,2\!-\!2\epsilon,3/2-\!\epsilon,3/2-\!\epsilon \\
             3/2,5/2-\!\epsilon,5/2-\!\epsilon \\
             \end{array};1
             \right]
\end{equation}
Using (\ref{eq:omega2}) its analytic continuation reads
\begin{align}
&\rho_7(\epsilon)=-\frac{\pi ^{5/2} 2^{2 \epsilon +1} \left(1-2\epsilon \right)^2 \csc (2 \pi  \epsilon ) \Gamma (-1 + 4 \epsilon)}{(1-2 \epsilon ) \Gamma (1-\epsilon ) \Gamma \left(\frac{3}{2}-\epsilon \right) \Gamma (2 \epsilon ) \Gamma^2 \left(-\frac{1}{2} +2 \epsilon \right)}\times\nonumber\\
&\times\sum\limits_{n=0}^{\infty}
\frac{\Gamma \left(n-\epsilon +\frac{1}{2}\right) \Gamma \left(n+2 \epsilon -\frac{1}{2}\right) \Gamma (n+2 \epsilon )}{n! \, \Gamma \left(n+\frac{3}{2}\right) \G\left(n+3 \epsilon -\frac{1}{2}\right)}
\left(\psi ^{(0)}\left(n+3 \epsilon +\tfrac{1}{2}\right)-\psi ^{(0)}\left(n-\epsilon +\tfrac{3}{2}\right)\right)
\end{align}
and for its $\epsilon$--expansion we obtain
\begin{align}
\rho_7(\epsilon)=&
\frac{1}{2 \epsilon ^2}+\frac{1}{\epsilon } \left(-\frac{\pi ^2}{2} + 1 + 4\log 2 \right)
\nonumber\\
&+\left(2-7 \zeta (3)+16 \log^2 2 + 8\log 2 - \frac{2}{3} \pi ^2 (3\log 2 -1)\right)+\mathcal{O}\left(\epsilon \right)
\end{align}

\vskip 8pt
We can now perform the sum of all these contributions as they appear in (\ref{intfcontr}) in order to obtain the result for $C_f$. Note that the individual pieces diverge with up to $\epsilon^{-3}$ poles and contain transcendental constants like $\zeta(3)$ and higher powers of $\log 2$. Remarkably, all these terms cancel in the final result, and it simply reads  
\begin{equation}
\boxed{ C_f = 4\pi^3 \left(\frac{1}{\epsilon} +2 - \gamma_E + 2 \log 2 - \log \pi \right)} + {\cal O}(\epsilon) 
\end{equation}
The result is still divergent, but when combined with the uncontracted part $U_f$ in eq. (\ref{eq:uncontracted}) the poles cancel and we obtain 
\begin{equation}
C_f+U_f = 32\pi^3 + {\cal O}(\epsilon) 
\end{equation} 
Therefore, multiplying by the overall factor $-\frac{1}{16\pi}\frac{MN}{k^2}$ in (\ref{CU}) we finally have
\begin{equation}\label{eq:resultf}
\boxed{\boxed{ {\rm (f)} = - 2 \pi^2\frac{M N}{k^2} }} + {\cal O}(\epsilon) 
\end{equation}
$~$
\vskip 15pt

\section{Final result and comparison with results from localization}
\label{localization}

We can now add contributions (\ref{WLoneloop}), (\ref{eq:bosonicresult}), (\ref{eq:Ieexpanded}) and (\ref{eq:resultf})
to obtain the expectation value of the 1/2 BPS WL up to two loops and for any value of $M$ and $N$
\begin{equation}\label{eq:result}
\boxed{\boxed{\left\langle W_{1/2} [\Gamma ] \right\rangle_{\text{f}=0} = 1 - \frac{\pi^2}{6\, k^2} \left( N^2+M^2 - 4\, NM - 1 \right) }}
\end{equation}
Here, the subscript refers to the fact that, as we review below, the perturbative evaluation in dimensional regularization corresponds to choosing framing zero. 

The finiteness of the result is a consequence of the fact that the contributing diagrams are separately finite. In fact, short distance divergences arise only in diagram 2(f) 
and at an intermediate stage, while its total contribution is eventually finite.

It is interesting to compare the result we have obtained with the exact result for the expectation value of 1/2 BPS WL derived using localization 
techniques. Let us first review briefly the latter.

Localization allows to reduce the computation of the partition function of a supersymmetric gauge theory on a sphere to the evaluation of a matrix model \cite{Pestun:2007rz}.
Similarly, correlation functions and WL can be computed by matrix model methods, provided the operators involved in the correlators are invariant under the same supercharge used for localizing the functional integral. If the matrix model is simple enough, a closed exact expression can be given for the expectation value of WL at all values of the coupling constant. Otherwise, a perturbative expansion is possible in some regions of the parameters, such as at small and large values of the coupling.
This provides results interpolating from weak to strong coupling, which can be used as tests of the AdS/CFT correspondence for theories that allow for a dual gravity description.

For supersymmetric CS theories and in particular ABJ(M) models on $S^3$ the partition function was found to be equivalent to the non--Gaussian matrix model \cite{Kapustin}
\begin{eqnarray}
\label{matrix}
\mathcal{Z}&=&\int \prod_{a=1}^{N}d\lambda _{a} \ e^{i\pi k\lambda
_{a}^{2}}\prod_{b=1}^{M}d\hat{\lambda }_{b} \ e^{-i\pi k\widehat{%
\lambda }_{b}^{2}} \times \label{Z}\\
&& \frac{\prod_{a<b}^{N}\sinh ^{2}(\pi (\lambda
_{a}-\lambda _{b}))\prod_{a<b}^{M}\sinh ^{2}(\pi (\hat{\lambda 
}_{a}-\hat{\lambda }_{b}))}{\prod_{a=1}^{N}\prod_{b=1}^{M}\cosh ^{2}(\pi (\lambda _{a}-\hat{\lambda }_{b}))}\nonumber 
\end{eqnarray}
Since $W_{1/6}[\Gamma ]$ and $W_{1/2}[\Gamma ]$ are invariant under the supercharge used to localize the functional integral \cite{Kapustin}, their expectation values can be computed by matrix model techniques. In particular, in \cite{DrukkerTrancanelli} a relation was proven connecting the 1/6 BPS and the 1/2 BPS circular WL.
Indeed, it was found that they belong to the same cohomology class with respect to the supercharge $Q$ used for localization. As a consequence, the evaluations of the corresponding expectation values via localization turn out to be equivalent. 

For computing 1/6 BPS WL we need insert in (\ref{matrix}) the factors 
\begin{equation}
w_{1/6}=\frac{1}{N}\sum_{a=1}^{N}e^{2\pi \lambda _{a}}\quad \text{%
and}\quad \hat{w}_{1/6}=\frac{1}{M}\sum_{a=1}^{M}e^{2\pi 
\hat{\lambda }_{a}}  \label{Wsum}
\end{equation}
which correspond to the $U(N)$ and $U(M)$ groups, respectively.  Instead, for the 1/2 BPS WL we need insert  
\beq
\label{sumW}
w_{1/2}=\frac{1}{N+M}\left( \sum_{a=1}^{N}e^{2\pi \lambda _{a}}+\sum_{a=1}^{M}e^{2\pi \hat{\lambda }_{a}} \right) =  \frac{N\,w_{1/6}+M\,\widehat{w}_{1/6}}{(N+M)}
\eeq

In ordinary perturbation theory, the evaluation of WL can in principle be hampered by singularities emerging from regions where two or more connections coincide on the circular path $\Gamma $.
One possible regularization has been investigated in knot theory where the problem of defining a topologically invariant regularization of a knot self--linking number has been addressed.
The regularization proposed in this scheme is known as framing procedure \cite{Witten, GMM, Labastida}.
It consists in deforming the original path $\Gamma $ into a nearby contour $\Gamma '$ by the introduction of a normal vector field along the contour, so allowing for a point splitting regularization of the correlation function. This procedure has been shown to provide sensible results for the pure non--abelian Chern--Simons theory \cite{GMM, Labastida}. It turns out that WL expectation values do not depend on the particular choice of the framing contour, but only on its topological properties with respect to the original path.
Wounding the framing contour on the original path $\text{f}$ times corresponds to framing $\text{f}$, and  $\big\langle W[\Gamma ]\big\rangle_{\text{f}}$ indicates the corresponding expectation value. It was shown that in this pure Chern-Simons context, different choices of framing will affect the result for $\big\langle W[\Gamma ]\big\rangle$ simply by an overall phase factor depending on $\text{f}$. 

Field theory computations are usually performed using alternative regularization schemes, so without framing (hereafter referred to as framing zero).
This entails great simplifications since many diagrams involve contractions of the $\varepsilon$ tensor with three vectors on the WL contour. Since in this case the contour is a circle that lies in a plane, such products vanish by antisymmetry, whereas if any of the vectors had been displaced from the plane, such as in the framing procedure, these terms wouldn't have vanished any longer.

In order for the framing procedure to be compatible with localization, it has 
to respect supersymmetry. This requires the original and the framing contours, $\Gamma $ and $\Gamma ' $, to be two fibers in the Hopf fibration of the $S^3$. Such great circles are linked once, hence the WL expectation value computed by localization corresponds to framing one \cite{Kapustin}. 
Therefore, in order to make a comparison with field theoretical computations one has to identify and remove the framing phase.

For the 1/6 BPS WL, the expectation values coming from localization expanded at weak coupling up to second order read \cite{Kapustin, Drukker:2010nc}
\begin{align}
\label{framing0}
& \big\langle W_{1/6}[\Gamma ]\big\rangle  _{\text{f}=1} = e^{\frac{i\pi}{k}N} \, \left[ 1 + \frac{\pi^2}{6\, k^2} \left( - N^2 + 6 MN + 1 \right) \right] \notag \\
& \big\langle \hat{W}_{1/6}[\Gamma ]\big\rangle  _{\text{f}=1} = e^{-\frac{i\pi}{k}M} \, \left[ 1 + \frac{\pi^2}{6\, k^2} \left( -M^2 + 6 MN + 1 \right) \right] 
\end{align}
Analogously, summing the contributions and normalizing as in (\ref{eq:Wplus}) we obtain the expectation value of $W^+_{1/6}$ in ABJ(M) at framing one  
\beq
\label{eq:Wplus2}
\langle W^+_{1/6} [ \G]\rangle_{\text{f}=1}= \frac{ N \langle W_{1/6} [ \G] \rangle_{\text{f}=1} + M \langle  \hat{W}_{1/6} [ \G] \rangle_{\text{f}=1}}{N+M}
\eeq
The overall phases appearing in (\ref{framing0}) are precisely those due to framing one. Consistently, the expectation values at framing zero are obtained removing such phases. In particular, if we remove these phases separately in the two terms appearing in (\ref{eq:Wplus2}) and sum the two contributions at framing zero we find perfect agreement with the perturbative field theory computation (\ref{eq:bosonicresult}). This includes the color subleading term also, which had not been considered before. We note that at this order subleading terms are framing independent, in the sense that they are not affected by a change of framing.

We now consider computing the 1/2 BPS operator in terms of the matrix model. This amounts to plugging the operator (\ref{sumW}) into the matrix model (\ref{Z}). Thanks to the algebraic identity appearing on the right hand side of eq. (\ref{sumW}), it is easy to realize that the expectation values, corresponding to framing one, are related by
\begin{equation}
\left\langle W_{1/2}[\Gamma ]\right\rangle _{\text{f}=1} =\frac{N\big\langle
W_{1/6}[\Gamma ]\big\rangle _{\text{f}=1} +M\big\langle \hat{W}_{1/6}[\Gamma ]\big\rangle _{\text{f}=1} }{N+M}
\label{eq:W}
\end{equation}
Therefore, the 1/2 BPS WL expectation value can be easily inferred from the 1/6 BPS WL results (\ref{framing0}) and coincides with $\langle W^+_{1/6} [ \G]\rangle$ at framing one, eq. (\ref{eq:Wplus2}). 

Plugging (\ref{framing0}) into the previous equation, we obtain
\begin{align}
\label{onehalfWL}
& \left\langle W_{1/2}[\Gamma ]\right\rangle _{\text{f}=1} = 1+\frac{i\pi }{k}(N-M) \notag \\
&~~~~~~~~~~ -\frac{\pi ^{2}}{6k^{2}}(4N^{2}+4M^{2}-10NM-1)+ 
\mathcal{O}(1/k^{3})
\end{align}
In order to compare this expression with the field theory computation we have to identify and remove the framing--one factor. While for $\langle W^+_{1/6} [ \G]\rangle$ the correct prescription is to remove the phase factors in the two terms of the linear combination (\ref{eq:Wplus2}) separately, for the 1/2 BPS WL the framing factor has been identified in \cite{Drukker:2010nc} as 
\begin{align}
\label{onehalfWL2}
\left\langle W_{1/2}[\Gamma ]\right\rangle _{\text{f}=1} &= e^{\frac{i\pi }{k}(N-M)} \left\langle W_{1/2}[C]\right\rangle _{\text{f}=0}    
\notag \\
& = e^{\frac{i\pi }{k}(N-M)} \,  \left[ 1 -\frac{\pi^2}{6\, k^2} \left( N^2+M^2 - 4 NM - 1 \right)+ \mathcal{O}(1/k^3) \right]
\end{align}
In particular, no remotion is required for $M=N$. 

The most important observation is that the expression within square brackets perfectly agrees with our field theory result (\ref{eq:result}) for any value of $M,N$. The perfect matching confirms that identification (\ref{onehalfWL2}) of the framing factor in the localization result is indeed the correct one. \\

\section{Conclusions}

For $U(N) \times U(M)$ ABJ(M) models we have computed analytically the vacuum expectation value of the 1/2 BPS circular Wilson loop in perturbation theory, up to two loops and for any value of $M,N$. 
Three years later, our result fills the gap between localization and perturbative calculations that had been left open since Drukker and Trancanelli made their prediction in \cite{DrukkerTrancanelli}. In fact, the perturbative two--loop result coincides with the weak coupling limit of the localization result, not only in the planar limit, but also when finite $M,N$ contributions are taken into account. 

In order to match the result from localization with the perturbative one, one needs to identify the non--trivial framing factor appearing in the localization expression and remove it. We have verified that the perturbative result matches the prediction from localization if we remove an overall phase as in eq. (\ref{onehalfWL2}), in agreement with the proposal of \cite{DrukkerTrancanelli}. Therefore, our calculation is a non--trivial confirmation of that proposal. 

The perturbative calculation involves a quite considerable number of integrals that require the development of sophisticated techniques to be solved. This is the reason why it has required few years for the problem to be tackled and solved. 

We have developed non--trivial strategies to overcome difficulties related to the appearance of divergent integrals, complicated parametric integrals and the necessity of analytic continuations to the physical region of parameters.  

We have handled short distance divergences by using dimensional regularization with dimensional reduction. In three dimensions and for Chern--Simons theories this is complicated by the appearance of the Levi--Civita tensor $\varepsilon_{\mu\nu\rho}$ that does not allow for any extension to dimensions different from three. We have circumvented this difficulty by performing tensorial algebra strictly in three dimensions and with a careful use of algebraic identities up to the point in which $\varepsilon$--tensors were no more present  in the loop structures. Only at that stage we have extended divergent integrals to $D = 3 -2\e$ dimensions. As a result of applying this procedure non--trivial evanescent factors arise, which multiply divergent integrals. These evanescent terms are crucial in determining the final result since they produce finite contributions when hit $\e$--poles. 

We have evaluated the integrals analytically in regions of the regularization parameter that make them well-defined. The main technical tools that we used are series expansions and  Mellin--Barnes representation.  
Most of the results turn out to be given in terms of hypergeometric functions that converge for (complex) dimensions that do not include the physical dimensions, $D=3$. Therefore, before taking the $\e \to 0$ limit suitable analytic continuation to regions that include a neighborhood of the origin is required. The non--trivial techniques to perform such a continuation have been detailed in Section 8. After analytic continuation the hypergeometric series can be expanded in powers of $\e$. Divergent pole contributions that appear at an intermediate stage of the expansion cancel out in the final result, which turns out to be finite, as expected. 

The procedure introduced in this paper allows in general to compute integrals appearing in circular Wilson loops at all orders in the regularization parameter.   
In principle, it could be used to evaluate more general Wilson loop operators \cite{Cardinali} and for more general theories.   \\ \\

\section*{Acknowledgements}

We thank L. Griguolo, M. Grisaru, J. Maldacena, and D. Seminara for very useful discussions.  
The work of MB has been supported by the Volkswagen-Foundation.
The work of GG and ML has been supported by the research project CONICET PIP0396.
The work of SP has been supported in part by INFN, MIUR--PRIN contract 2009--KHZKRX and MPNS--COST Action MP1210 "The String Theory Universe".

\vfill
\newpage

\appendix

\section{Conventions and Feynman rules}\label{sec:conventions}

We work in euclidean three--dimensional space with coordinates $x^\mu = (x^0, x^1, x^2)$. The conventions are obtained by Wick rotating the ones in Minkowski \cite{Drukker}. 

Applying the prescription  $\g^0_E = -i \g^0$, the euclidean gamma matrices satisfying Clifford algebra $\{ \g^\mu , \g^\nu \} = 2 \d^{\mu\nu} \mathbb{I}$, are defined as
\beq
(\g^\mu)_\a^{\; \, \b} = \{ -\s^3, \s^1, \s^2 \}
\eeq
with matrix product 
\beq
\label{prod}
(\g^\mu \g^\nu)_\a^{\; \, \b} \equiv (\g^\mu)_\a^{\; \, \g} (\g^\nu)_\g^{\; \, \b}
\eeq
Useful identities are
\bea 
\label{id1}
&&  \g^\mu \g^\nu = \d^{\mu \nu} \mathbb{I} - i \varepsilon^{\mu\nu\rho} \g^\rho
\non \\
&& \g^\mu \g^\nu \g^\rho = \d^{\mu\nu} \g^\rho - \d^{\mu\rho} \g^\nu+  \d^{\nu\rho} \g^\mu  - i \varepsilon^{\mu\nu\rho} \mathbb{I}
\non \\
&& 
\g^\mu \g^\nu \g^\rho \g^\s -  \g^\s \g^\rho \g^\nu \g^\mu = -2i \left( \d^{\mu\nu} \varepsilon^{\rho\s \eta}  + \d^{\rho \s}  \varepsilon^{\mu\nu\eta} + \d^{\nu\eta} \varepsilon^{\rho \mu \s} +
\d^{\mu\eta} \varepsilon^{\nu\rho\s}  \right) \g^\eta
 \\
&& 
\non \\
&&
\Tr (\g^\mu \g^\nu) = 2 \d^{\mu\nu}
\non \\
&&
\Tr (\g^\mu \g^\nu \g^\rho) = -2i \varepsilon^{\mu\nu\rho}
\eea 
Spinorial indices are lowered and raised as $(\g^\mu)^\a_{\; \, \b} = \varepsilon^{\a \g}  (\g^\mu)_\g^{\; \, \d} \varepsilon_{\b \d}$, where
\beq
\varepsilon^{\a\b} =  \left( \begin{array}{cc} 0 & 1 \\ -1 & 0 \end{array} \right) 
\qquad \qquad 
\varepsilon_{\a\b} =  \left( \begin{array}{cc} 0 & -1 \\ 1 & 0 \end{array} \right) 
\eeq
It follows that 
\beq
(\g^\mu)^\a_{\; \, \b} = \{ - \s^3, \s^1, -\s^2 \} = (\g^\mu)^T
\eeq
In addition,
\bea
&& (\g^\mu)_{\a \b} = \{ - \s^1, -\s^3, i \mathbb{I} \} = (\g^\mu)_{\b\a}  
\non \\
&& (\g^\mu)^{\a \b} = \{ \s^1, \s^3, i \mathbb{I} \} = (\g^\mu)^{\b\a}  
\eea
are symmetric matrices.

We conventionally choose the spinorial indices of chiral fermions to be always up, while the ones of antichirals to be always down. Therefore, the spinorial product is always meant to be
\bea
&& \psi_1 \bar{\psi}_2 \equiv \psi_1^\a \bar{\psi}_{2 \a} = - \bar{\psi}_{2 \a}  \psi_1^\a \equiv - \bar{\psi}_2 \psi_1 \qquad {\rm (for ~anticommuting ~spinors)}
\non \\
&& \psi \bar{\eta} \equiv \psi^\a \bar{\eta}_\a =  \bar{\eta}_\a  \psi^\a \equiv \bar{\eta} \psi \qquad \qquad \qquad ~ {\rm (for ~commuting ~spinors)}
\eea~
With this convention we write
\beq
\bar{\eta}_1 \g^\mu \eta_2 \equiv \bar{\eta}_{1 \; \a}  (\g^\mu)^\a_{\; \, \b}  \eta_2^\b = \eta_2^\b  (\g^\mu)^{\; \, \a}_\b \bar{\eta}_{1 \; \a}
\equiv \eta_2 \g^\mu \bar{\eta}_1 
\eeq
Moreover, in the text we indicate
\beq
(\eta_1 \g^\mu \bar{\eta}_2) \equiv \eta_{1 I} \g^\mu \bar{\eta}^I_2
\eeq
where a sum over the $SU(4)_R$ index is understood. 

The ${\cal N} = 6$ supersymmetric Chern--Simons--matter theory \cite{ABJM, ABJ} with gauge group $U(N) \times U(M)$ is described by the euclidean action ($\G = \int e^{-S}$)
\beq
S = S_{CS} + S_{matter} + S_{gf} 
\eeq
\bea
\label{action}
S_{CS} &=& -i \frac{k}{4\pi}\int d^3x\,\varepsilon^{\mu\nu\rho} \Big[ \Tr \left( A_\mu\partial_\nu A_\rho+\frac{2}{3} i A_\mu A_\nu A_\rho \right)
 \\
&~& \qquad \qquad \qquad \qquad \quad - \Tr \left(\hat{A}_\mu\partial_\nu 
\hat{A}_\rho+\frac{2}{3} i \hat{A}_\mu \hat{A}_\nu \hat{A}_\rho \right) 
\Big]
\non \\
S_{matter} &=& \int d^3x \, \Tr \Big[ D_\mu C_I D^\mu \bar{C}^I + i \bar{\psi}^I \g^\mu D_\mu \psi_I \Big] + S_{int} 
\non \\
S_{gf} &=& \frac{k}{4\pi} \int d^3x \, \Tr \Big[ \frac{1}{\xi}  (\pa_\mu A^\mu)^2 + \pa_\mu \bar{c} D^\mu c  - 
\frac{1}{\xi} ( \pa_\mu \hat{A}^\mu )^2 - \pa_\mu \bar{\hat{c}} D^\mu \hat{c} \Big] \non
\eea
where $(C_I)^j_{\; \hat{j}}$ ($(\bar{C}^I)^{\hat{j}}_{\; j}$), $I=1, \cdots  4$,  are  four matter scalars in the  bifundamental (antibifundamental) representation of the gauge group, 
whereas $(\bar{\psi}^I)^j_{\; \hat{j}}$ ($(\psi_I)^{\hat{j}}_{\; j}$) are the corresponding fermions. 

The covariant derivatives are  defined as
\bea
\label{covariant}
D_\mu C_I &=& \pa_\mu C_I + i A_\mu C_I - i C_I \hat{A}_\mu
\non \\
D_\mu \bar{C}^I &=& \pa_\mu \bar{C}^I - i \bar{C}^I A_\mu + i \hat{A}_\mu \bar{C}^I  
\non \\
D_\mu \bar{\psi}^I  &=& \pa_\mu \bar{\psi}^I + i A_\mu \bar{\psi}^I - i \bar{\psi}^I \hat{A}_\mu
\non \\
D_\mu \psi_I &=& \pa_\mu \psi_I - i \psi_I A_\mu + i \hat{A}_\mu \psi_I  
\eea
With these assignments the action is invariant under the following gauge transformations
\bea
\label{gauge}
&& A'= U A\, U^\dagger- i\, U d U^\dagger \quad ~, \quad \hat{A}'= \hat{U} \hat{A}\, \hat{U}^\dagger- i\, \hat{U} d \hat{U}^\dagger 
\non \\
&& 
\quad \quad \phi' = U \phi \hat{U}^\dagger  \qquad \qquad , \qquad  \quad \bar{\phi}' = \hat{U} \bar{\phi} U^\dagger
\eea
where $U$and $\hat{U}$ are the transformation matrices for the groups $U(N)$ and $U(M)$ respectively, and $\phi$ ($\bar{\phi}$) stands for any field in the (anti)bifundamental.  

The Wilson loop we are interested in is defined as
\beq
\label{WL2}
W_{1/2}[\G] = \frac{1}{N+M} \Tr \,P \exp{ \left( -i \int_\G d\tau {\cal L}(\tau)\right) } 
\eeq 
with ${\cal L}$ given in eq. (\ref{supermatrix}).  We stress that the sign in front of the integral is unambiguously fixed by gauge invariance under transformations (\ref{gauge}).  

From the action (\ref{action}) we obtain the following Feynman rules: 

\vskip 15pt
\noindent
\underline{The propagators} 
\begin{itemize}
\item Tree--level vector propagators in Landau gauge
\bea
\label{treevector}
&& \langle A_\mu^a (x) A_\nu^b(y) \rangle^{(0)} =  \d^{ab}   \, \left( \frac{2\pi i}{k} \right) \frac{\G(\frac32-\e)}{2\pi^{\frac32 -\e}} \varepsilon_{\mu\nu\rho} \frac{(x-y)^\rho}{[(x-y)^2]^{\frac32 -\e} }
\non \\
&& \langle \hat{A}_\mu^a (x) \hat{A}_\nu^b(y) \rangle^{(0)} =  -\d^{ab}   \, \left( \frac{2\pi i}{k} \right) \frac{\G(\frac32-\e)}{2\pi^{\frac32 -\e}} \varepsilon_{\mu\nu\rho} \frac{(x-y)^\rho}{[(x-y)^2]^{\frac32 -\e} }
\eea
\item One--loop vector propagators (see for instance \cite{Griguolo})
\bea
\label{1vector}
&& \langle A_\mu^a (x) A_\nu^b(y) \rangle^{(1)} = \d^{ab}   \left( \frac{2\pi }{k} \right)^2 N \frac{\G^2(\frac12-\e)}{4\pi^{3 -2\e}} 
\left[ \frac{\d_{\mu\nu}}{ [(x- y)^2]^{1-2\e}} - \pa_\mu \pa_\nu \frac{[(x-y)^2]^\e}{4\e(1+2\e)} \right]  \non  \\
&& \langle \hat{A}_\mu^a (x) \hat{A}_\nu^b(y) \rangle^{(1)} = \d^{ab}   \left( \frac{2\pi }{k} \right)^2 M \frac{\G^2(\frac12-\e)}{4\pi^{3 -2\e}} 
\left[ \frac{\d_{\mu\nu}}{ [(x- y)^2]^{1-2\e}} - \pa_\mu \pa_\nu \frac{[(x-y)^2]^\e}{4\e(1+2\e)} \right] \non \\
\eea
\item Scalar propagator
\beq
\label{scalar}
\langle (C_I)_i^{\; \hat{j}} (x) (\bar{C}^J)_{\hat{k}}^l(\; y) \rangle^{(0)}  = \d_I^J \d_i^l \d_{\hat{k}}^{\hat{j}} \, \frac{\G(\frac12 -\e)}{4\pi^{\frac32-\e}} 
\, \frac{1}{[(x-y)^2]^{\frac12 -\e}}
\eeq
\item Tree--level fermion propagator
\beq
\label{treefermion}
\langle (\psi_I^\a)_{\hat{i}}^{\; j}  (x) (\bar{\psi}^J_\b )_k^{\; \hat{l}}(y) \rangle^{(0)} = - i \, \d_I^J \d_{\hat{i}}^{\hat{l}} \d_{k}^{j} \, 
\frac{\G(\frac32 - \e)}{2\pi^{\frac32 -\e}} \,  \frac{(\g^\mu)^\a_{\; \, \b} \,  (x-y)_\mu}{[(x-y)^2]^{\frac32 - \e}}
\eeq
\item One--loop fermion propagator \cite{Griguolo}
\beq
\label{1fermion}
\langle (\psi_I^\a)_{\hat{i}}^{\; j}  (x) (\bar{\psi}^J_\b)_k^{\; \hat{l}}(y) \rangle^{(1)} = - i \,\left( \frac{2\pi}{k} \right) \,  \d_I^J \d_{\hat{i}}^{\hat{l}} \d_{k}^{j} \,  \, \d^\a_{\; \, \b}
\, (M-N) \frac{\G^2(\frac12 - \e)}{16 \pi^{3-2\e}} \, \frac{1}{[(x-y)^2]^{1 - 2\e}}  
\eeq	 
\end{itemize}	 

\vskip 15pt
\noindent
\underline{The interaction vertices}
\begin{itemize}
\item Gauge cubic vertex
\beq
\label{gaugecubic}
-i \frac{k}{12\pi} \varepsilon^{\mu\nu\rho} \int d^3x \, f^{abc} A_\mu^a A_\nu^b A_\rho^c
\eeq
\item Gauge--fermion cubic vertex
\beq
\label{gaugefermion}
-\int d^3x \, \Tr \Big[ \bar{\psi}^I \g^\mu \psi_I A_\mu - \bar{\psi}^I \g^\mu \hat{A}_\mu \psi_I  \Big]
\eeq 
\end{itemize}

Finally, we recall our color conventions. We work with hermitian generators for $U(N)$ and $U(M)$ gauge groups, satisfying
\beq 
\Tr (T^a T^b ) = \d^{ab} \qquad , \qquad \Tr (\hat{T}^{\hat{a}} \hat{T}^{\hat{b}} ) = \d^{\hat{a} \hat{b}}
\eeq
and
\bea
\sum_{a=1}^{N^2} (T^a)_{ij} (T^a)_{kl} = \d_{il} \d_{jk}  \qquad &,& \qquad \sum_{\hat{a}=1}^{M^2} (\hat{T}^{\hat{a}})_{ij} (\hat{T}^{\hat{a}})_{kl} = \d_{il} \d_{jk}
\non \\
f^{abc} f^{abc} = 2 N^3 \qquad &,& \qquad  f^{\hat{a}\hat{b}\hat{c}} f^{\hat{a}\hat{b}\hat{c}} = 2 M^3 
\eea

\section{Useful identities on the unit circle}\label{sec:formulitas}

We parametrize a point on the unit circle $\G$ as 
\beq
x_i^\mu = (0, \cos{\tau_i}, \sin{\tau_i}) \quad , \quad \dot{x}_i^\mu = (0, -\sin{\tau_i}, \cos{\tau_i}) \quad , \quad | x_i|^2 = 1
\eeq 
Simple identities that turn out to be useful along the calculation are
\bea
\label{I1}
&& (x_i - x_j)^2 = 4 \sin^2{\frac{\tau_i-\tau_j}{2}}
\\
\label{I2}
&& x_i \cdot x_j = \dot{x}_i \cdot \dot{x}_j = \cos{(\tau_i - \tau_j)}
\\
\label{I3}
&& x_i \cdot \dot{x}_j = \sin{(\tau_i - \tau_j)}
\\
\label{I4}
&& (x_i \cdot x_j) (\dot{x}_i \cdot \dot{x}_j)  - (x_i \cdot \dot{x}_j ) (\dot{x}_i \cdot x_j ) = 1
\\
\label{I5}
&& (x_i-x_j) \cdot  (\dot{x}_i + \dot{x}_j) = 2 \sin{(\tau_i - \tau_j)}
\eea

Using expression (\ref{DT}) for the $\eta$ spinors and writing $\eta_i \equiv \eta(\tau_i)$, ${\mathcal M}_i  \equiv {\mathcal M}(\tau_i)$ a list of useful identities follows
\bea
&& (\eta_{i} \bar \eta_{j}) = 2i \cos{\frac{\tau_i-\tau_j}{2}}
 \label{eq:etaetabar}   \\
&& (\eta_{i} \gamma_0 \bar\eta_{j}) = 2 \sin{\frac{\tau_i-\tau_j}{2}}
\label{eq:etagamma0}  \\
&& (\eta_{i} \gamma_1 \bar\eta_{j}) =  - 2i \sin{\frac{\tau_i + \tau_j}{2}}
\\
&& (\eta_{i} \gamma_2 \bar\eta_{j}) = 2i \cos{\frac{\tau_i + \tau_j}{2}} 
\\
&& \Tr({\mathcal M}_i  {\mathcal M}_j) =  4
\label{eq:matrices} \\
&& (\eta_{i} \gamma_{\mu} \bar \eta_{j})\, (x_i - x_j)^{\mu} =  4i \, \sin{\frac{\tau_i - \tau_j}{2}}  
\label{eq:etagammax}
\eea
More generally, we can write
\beq
\label{eq:etagammaeta}
(\eta_{i} \gamma^\mu \bar \eta_{j}) = -\frac{2}{(\eta_{i} \bar\eta_{j})} \Big[ \dot{x}_i^{\mu} + \dot{x}_j^{\mu} + i \, \varepsilon_ {\; \; \rho \nu}^{ \mu} \, \dot{x}_i^\rho \, \dot{x}_j^\nu \Big]
\eeq

\section{Method for solving circle integrals}\label{sec:method}

In this Appendix we spell out the method we employ to solve complicated trigonometric multiple integrals as the ones arising from diagrams 2(e), 2(f).  

Given a multiple integral of the product of sine and cosine functions to some power, the starting point consists in replacing each trigonometric function by its complex exponential form 
\begin{align}
&\sin^{-\a} f(\{\t\}) \rightarrow (2i)^{\a}\, \left( e^{i f(\{\t\})} - e^{-i f(\{\t\})} \right)^{-\a}
\nonumber\\&
\cos^{-\a} f(\{\t\}) \rightarrow 2^{\a}\, \left( e^{i f(\{\t\})} + e^{-i f(\{\t\})} \right)^{-\a}
\end{align} 
In the integrals we have to evaluate, $f$ is a real linear function of the $\{ \tau \}$ parameters, and $\a$ is a complex parameter typically linear in the dimensional regularization parameter $\e$.

Next we expand each factor, obtaining  
\begin{align}
\label{expansion}
& \sin^{-\a} f(\{\t\}) \rightarrow (2i)^{\a}\, \frac{1}{\Gamma (\a)}\, \sum_{n=0}^{\infty}\, \frac{\Gamma (n+\a)}{n!}\, \left(e^{-i f(\{\t\})}\right)^{2n+\a}
\nonumber\\&
\cos^{-\a} f(\{\t\}) \rightarrow 2^{\a}\, \frac{1}{\Gamma (\a)}\, \, \sum_{n=0}^{\infty}\, (-1)^n\, \frac{\Gamma (n+\a)}{n!}\, \left(e^{-i f(\{\t\})}\right)^{2n+\a}
\end{align}
If we exchange the parametric integral with the series, the integrals can be easily performed.

Some comments on the mathematical consistency of these steps are in order. It should be mentioned that since $f(\{\t\})$ is a real function, then the power expansions (\ref{expansion}) is made in terms of a unimodular variable. 
{This means that the series converges absolutely only for $\Re(\alpha)<0$. However, being $\alpha$ a function of $\epsilon$, we can perform the calculation in regions of values for $\epsilon$   where the series converges absolutely and then analytically continue the result to all $\epsilon$.

We proceed with the evaluation of each single integral. This produces multiple series, which we eventually sum in terms of hypergeometric functions.}
A posteriori, we check the consistency of our procedure by performing an exhaustive numerical comparison between the result obtained and the original integral for a sufficiently large range of complex values of  $\epsilon$  where the integral converges.

As an example, we solve the following parametric integral
\begin{equation}\label{eq:1loopint}
I^{(1)} = \int_{0}^{2\pi} d\t_1\, \int_{0}^{\t_1} d\t_2\; \frac{1}{\left[ \sin^{2}\frac{\t_{12}}{2} \right]^{\a}} 
\end{equation}
where $\a$ is a generic complex parameter. When $\a = (1 -\e)$ this is the integral that appears in the one--loop contribution to the 1/2 BPS WL (see Section 3).

We can use the parameter $\a$ as a regulator, in the spirit of dimensional regularization: We compute the integral in the domain where it converges, which is for $\Re(\a)<\tfrac12$ and then analytically extend it for any value of $\a$.

We rewrite the integral as
\begin{equation}
I^{(1)} =  (2i)^{2\alpha}\int\limits_{0}^{\,\,2\pi} d\t_1\int\limits_{0}^{\,\,\t_1}d\t_2
\frac{\left(e^{-i \t_{12}}\right)^{\alpha}}
{\left(1-e^{-i \t_{12}}\right)^{2\alpha}};
\end{equation}
Expanding the denominator and integrating term by term we have
\begin{equation}
I^{(1)} =  \frac{(2i)^{2\alpha}}{\Gamma(2\alpha)}\sum\limits_{n=0}^{\infty}
\frac{\Gamma(n+2\alpha)}{n!}
\int\limits_{0}^{\,\,2\pi} d\t_1\int\limits_{0}^{\,\,\t_1}d\t_2
\left(e^{-i \t_{12}}\right)^{\alpha+n}
\end{equation}
The integrals can be easily solved. Introducing  the shorthand notation
\begin{equation}
\label{Sseries}
S_\lambda[\alpha]=\sum\limits_{n=0}^{\infty}
\frac{\Gamma(n+2\alpha)}{n!(n+\alpha)^\lambda}
\end{equation}
we can write
\beq
\label{twosums}
I^{(1)}=\frac{4^{\alpha}}{\Gamma(2\alpha)}
\left(
-2\,i\pi e^{i\pi\alpha} S_1[\alpha]+2\,i\sin(\pi\alpha)\,S_2[\alpha]
\right)
\eeq
This is in general a complex function. We can further simplify its form by imposing it to be real for $\a$ real. 
Since for $\a$ real the two series sum to real functions, selecting the 
imaginary part of (\ref{twosums}) and setting it to zero leads to a non--trivial 
relation between the two series, $S_2[\alpha] = \pi \cot{\a} S_1[\a]$.
 In this simple case, this identity can be directly checked by comparing the two explicit summations
\bea\label{eq:oneloopseries}
S_1[\a] &=& \frac{2^{-2\alpha } \Gamma \left(\frac{1}{2}-\alpha \right) \Gamma (\alpha ) \Gamma (2 \alpha )}{\sqrt{\pi }}
\non \\
S_2[\a] &=& \sqrt{\pi }\, 2^{-2\alpha } \cot (\pi  \alpha ) \Gamma \left(\tfrac{1}{2}-\alpha \right) \Gamma (\alpha ) \Gamma (2 \alpha )
\eea
However, in more complicated cases where summing the series is not an easy task, the trick of imposing the reality of the result for $\a$ real turns out to be very convenient for deriving identities between series that might be difficult to prove otherwise.

Using the previous findings, we finally obtain
\beq
\label{onesum}
I^{(1)}=\frac{2^{2\alpha+1}\, \pi }{\Gamma(2\alpha)}
\sin(\pi\alpha) S_1[\alpha] = \frac{2 \pi ^{3/2} \Gamma \left(\frac{1}{2}-\alpha \right)}{\Gamma \left(1-\alpha \right)}
\eeq
As can be checked numerically, this is the correct result for any (even complex) value of $\a$. 

We note that for the one--loop contribution to the WL, setting $\a = (1 -\e)$ we indeed find a ${\cal O}(\e)$ result.  

As a second application, we prove that the sum of the three integrals in eqs. (\ref{planarintegrals}, \ref{crossedintegral}) is subleading in $\e$. In fact,  
summing the three contributions we are left with an integrand which is totally symmetric under exchanges of the integration variables. This allows to symmetrize the integration domain 
\begin{equation}
I_{\rm e}^{(1)}+I_{\rm e}^{(2)}+I_{\rm e}^{(3)} = 
\frac{3}{4!}\, \int_{0}^{2\pi} d\t_1 \, \int_{0}^{2\pi} d\t_2\,
\frac{1}{\left(\sin^2 \frac{\t_{12}}{2} \right)^{\a}}
\int_{0}^{2\pi} d\t_3\,\int_{0}^{2\pi} d\t_4\, 
\frac{1}{\left( \sin^2 \frac{\t_{34}}{2} \right)^{\a}}
\end{equation}
leading to a factorized expression in terms of one--loop integrals of the type (\ref{eq:1loopint}).
Using the general result (\ref{onesum}) we then obtain 
\begin{equation}
I_{\rm e}^{(1)}+I_{\rm e}^{(2)}+I_{\rm e}^{(3)} = 
\frac{2 \pi ^{3} \Gamma^2\left(- \frac{1}{2}+\e \right)}{\Gamma^2(\e)} 
\end{equation}
This expression is subleading in $\e$, as stated above.

\section{Reduction to a scalar integral}
\label{scalarintegral}

In this Appendix we prove that the linear combination of vertex integrals
\beq\label{needtoevaluate}
(\eta_2\gamma_\mu\bar\eta_3) \, \varepsilon_{\nu\rho\sigma}\dot{x}_1^\nu
\left(\Gamma^{\sigma\rho\mu}+\Gamma^{\sigma\mu\rho}\right)
\eeq
appearing in eq. (\ref{1fintegral}) can be reduced to a scalar triangle integral. 

Applying Feynman combining in $D=3-2\e$ dimensions,  every single vertex integral can be written as $\Gamma^{\mu\nu\rho}= \partial^{\mu}_1\partial^{\nu}_2\partial^{\rho}_3 { \mathcal G}$, where 
\begin{align}
{\mathcal G} &= \int d^{3-2\epsilon}x \,
\frac{1}
{\left[(x-x_1)^2\,(x-x_2)^2\,(x-x_3)^2\right]^{1/2-\epsilon}} = \nonumber\\& =
\frac{\pi^{3/2-\epsilon}\,\Gamma(-2\epsilon)}{\Gamma^3(\tfrac{1}{2}-\epsilon)}
\int [d\alpha]_3
\frac{(\alpha_1\alpha_2\alpha_3)^{-1/2-\epsilon}}
{\left(\Omega^2\right)^{-2\epsilon}}
\end{align}
Here the measure is defined as $[d\alpha]_3=\prod_{i=1}^3 d\alpha_i\,\delta(\sum_{i=1}^3 \alpha_i\!-\!1)$ and 
\beq
\Omega^2=\alpha_1\alpha_2 x_{12}^2+\alpha_2\alpha_3 x_{23}^2+\alpha_1\alpha_3 x_{13}^2
\eeq
When applying the derivatives to $\mathcal{G}$ we are interested only in terms proportional to the metric tensor, since terms proportional to the product of coordinate vectors would be zero for the planarity of the contour. Therefore, we can write
\begin{align}\label{E4}
\Gamma^{\mu\nu\rho} \rightarrow
& \frac{4\pi^{3/2-\epsilon}\,\Gamma(2-2\epsilon)}{\Gamma^3(\tfrac{1}{2}-\epsilon)}
\int [d\alpha]_3\,
\frac{(\alpha_1\alpha_2\alpha_3)^{1/2-\epsilon}}
{\left(\Omega^2\right)^{2-2\epsilon}}
\Big{[}
\hat{\eta}^{\mu\nu}(\alpha_1 x_{13}^\rho+\alpha_2 x_{23}^\rho)+
\nonumber\\
&
+\hat{\eta}^{\nu\rho}(\alpha_2 x_{21}^\mu+\alpha_3 x_{31}^\mu)+
\hat{\eta}^{\rho\mu}(\alpha_1 x_{12}^\nu+\alpha_3 x_{32}^\nu)
\Big{]} 
\end{align}
Introducing the notation 
\beq
\label{integralI}
I_{\bf i}=\frac{4\pi^{3/2-\epsilon}\,\Gamma(2-2\epsilon)}{\Gamma^3(\tfrac{1}{2}-\epsilon)}
\int [d\alpha]_3\,
\frac{(\alpha_1\alpha_2\alpha_3)^{1/2-\epsilon}\ \ \alpha_{\bf i}}
{\left(\Omega^2\right)^{2-2\epsilon}}
\eeq 
we rewrite the last expression as
\beq
\label{GammaMetric}
\Gamma^{\mu\nu\rho} \rightarrow
\hat{\eta}^{\mu\nu}(I_1 x_{13}^\rho+I_2 x_{23}^\rho)+
\hat{\eta}^{\nu\rho}(I_2 x_{21}^\mu+I_3 x_{31}^\mu)+
\hat{\eta}^{\rho\mu}(I_1 x_{12}^\nu+I_3 x_{32}^\nu) 
\eeq
and insert it in eq. (\ref{needtoevaluate}). 

We stress that the metric tensor appearing here is a $D$--dimensional metric, being it produced from the evaluation of a $D$--dimensional tensor integral. Caution is then required when contracting it with $\varepsilon_{\nu\rho\sigma}$. As discussed in the main text, a safe prescription is to get rid of products of Levi--Civita tensors  
in favour of three--dimensional metric tensors and then use identities (\ref{DRED}). 

To this end, we rewrite the spinorial structure $(\eta_2\gamma_\mu\bar{\eta}_3)$ in eq. (\ref{needtoevaluate}) with the help of identity (\ref{eq:etagammaeta}). It is easy to realize that the first two terms in this equation do not contribute in (\ref{needtoevaluate}) due to the planarity of the contour. From the third term, collecting everything, we obtain
\begin{align}
(\ref{needtoevaluate})=&
-\frac{2 i}{(\eta_2\bar{\eta}_3)}\,
\varepsilon_{\mu\alpha\beta}\,\varepsilon_{\nu\rho\sigma}\,
\dot{x}^\nu_1 \dot{x}^\alpha_2 \dot{x}^\beta_3\times\nonumber\\
&
\big{[}
\hat{\eta}^{\sigma\mu}(I_1 x_{12}^\rho+I_3 x_{32}^\rho+I_1 x_{13}^{\rho}+I_2 x_{23}^\rho)+
2\hat{\eta}^{\rho\mu}(I_2 x_{21}^\sigma+I_3 x_{31}^\sigma)
\big{]}
\end{align}
It is now easy to use identity (\ref{epsilon}) to trade the product of the two $\varepsilon$ tensors with products of three dimensional metric tensors. Applying the rules (\ref{DRED}) and using the explicit realization of $\eta$ spinors and the parametrization of the curve, after some work, we obtain
\begin{align}\label{E8}
& (\eta_2\gamma_\mu\bar\eta_3) \, \varepsilon_{\nu\rho\sigma}\dot{x}_1^\nu
\left(\Gamma^{\sigma\rho\mu}+\Gamma^{\sigma\mu\rho}\right)  = 
\frac{2i(D-2)}{(\eta_2\bar{\eta}_3)}
(\dot{x}_{2\rho}\,\dot{x}_1 \cdot \dot{x}_3-\dot{x}_{3\rho}\,\dot{x}_1 \cdot \dot{x}_2)
\left(x_{12}^{\rho}+x_{13}^{\rho}\right)\,I(x)
\non \\
&=   
-4 (D-2)\sin{\tfrac{\tau_{23}}{2}}
\left[
\sin^2{\tfrac{\tau_{12}}{2}}+
\sin^2{\tfrac{\tau_{13}}{2}}
\right] \, I(x)
\end{align}
where $\tau_{ij} = \tau_i - \tau_j$ and we have defined $I(x)\equiv I_1+I_2+I_3$ with $I_{\bf i}$ given in (\ref{integralI}). We stress that the appearance of an overall factor $(D-2)=(1-2\epsilon)$ is the result of applying the DRED prescription (\ref{DRED}).

Thanks to the condition $\alpha_1\!+\!\alpha_2\!+\!\alpha_3\!=\!1$ coming from the delta function inside the measure, the explicit expression of the $I(x)$ integral reads
\beq
I(x)=\frac{4\pi^{3/2-\epsilon}\,\Gamma(2-2\epsilon)}{\Gamma^3(\tfrac{1}{2}-\epsilon)}
\int [d\alpha]_3\,
\frac{(\alpha_1\alpha_2\alpha_3)^{1/2-\epsilon}}
{\left(\Omega^2\right)^{2-2\epsilon}}
\eeq
Performing a Mellin--Barnes transformation we obtain
\begin{align}\label{eq:MB}
I(x) = & \frac{(1-2\epsilon)^3}{2\pi}\, \frac{\pi^{\frac52 - \e}}{\Gamma^3\left( \frac32 - \e \right)\Gamma\left( \frac12 + \e \right)}\, \int \frac{du\, dv}{(2\pi i)^2} \Gamma\left( -u \right) \Gamma\left( -v \right) \Gamma\left( -u -\frac12 + \e \right) 
\nonumber\\ &
\Gamma\left( -v -\frac12 + \e \right) \Gamma\left( \frac32 -\e + u + v \right) \Gamma\left( 2 - 2\e+ u + v \right)
\frac{(x_{12}^2)^{u} (x_{23}^2)^{v}}{(x_{13}^2)^{u+v+2-2\e}}
\nonumber\\ &
\nonumber\\ = &
2^{-2+4\e}\, \frac{\pi^{\frac32 - \e}}{\Gamma^3\left( \frac12 - \e \right)\Gamma\left( \frac12 + \e \right)}\, \int \frac{du\, dv}{(2\pi i)^2} \Gamma\left( -u \right) \Gamma\left( -v \right) \Gamma\left( -u -\frac12 + \e \right) 
\nonumber\\ &
\Gamma\left( -v -\frac12 + \e \right) \Gamma\left( \frac32 -\e + u + v \right) \Gamma\left( 2 - 2\e+ u + v \right)
\frac{\left(\sin^2 \frac{\t_{12}}{2}\right)^u \left(\sin^{2} \frac{\t_{23}}{2}\right)^v}{\left(\sin^2 \frac{\t_{13}}{2}\right)^{u+v+2-2\e}}
\end{align}

\section{Evaluation of uncontracted integrals}
\label{app:uncontracted}

In this Appendix we give the details for the evaluation of the integral in eq. (\ref{U2}) arising as part of diagram \ref{2loop}(f). 

We first concentrate on the parametric ${\cal J}$--integral defined in eq. (\ref{J}).
Performing the shift of integration variables $\tau_3\rightarrow\tau_3+\tau_1$, $\tau_2\rightarrow\tau_2+\tau_1$ and exploiting the $2\pi$-periodicity of the integrand as a function of $(\tau_2, \tau_3)$ 
it can be rewritten as 
\begin{equation}
\mathcal{J}(\alpha,\beta,\gamma)=
2\pi\int _{0}^{2\pi}\!\!\!\!d\tau_2\int _{0}^{2\pi}\!\!\!\! d\tau_3
\left[\sin ^{2}\left(\frac{\tau _{2}}{2}\right)\right]^\alpha
\left[\sin ^{2}\left(\frac{\tau _{23}}{2}\right)\right]^\beta
\left[\sin ^{2}\left(\frac{\tau _{3}}{2}\right)\right]^\gamma
\end{equation}
where the $2\pi$ factor comes from the trivial integration over $\tau_1$. The remaining integrations can be carried out by using the procedure described in Appendix \ref{sec:method}. Writing the trigonometric functions in terms of exponentials, expanding them as series and integrating term by term we end up with triple series. Imposing reality of the result for $\alpha$, $\beta$ and $\gamma$ real, we obtain  
\begin{align}
& \mathcal{J}(\alpha,\beta,\gamma)= 
2\pi\, \sum_{l=0}^{\infty}\, \sum_{m=0}^{\infty}\, \sum_{n=0}^{\infty}\, \frac{2^{-2 (\alpha +\beta +\gamma )} \Gamma (l-2 \alpha ) \Gamma (m-2 \beta ) \Gamma (n-2 \gamma )}{\Gamma (-2 \alpha ) \Gamma (-2 \beta ) \Gamma (-2 \gamma ) \Gamma (l+1) \Gamma (m+1) \Gamma (n+1)}
\nonumber\\ &
~~~~ \left(
\frac{\cos (\pi  (\alpha -\beta -\gamma ))}{(-\alpha +\beta +l-m) (-\beta -\gamma +m+n)}
-\frac{\cos (\pi  (\alpha +\beta -\gamma ))}{(-\alpha -\beta +l+m) (-\beta +\gamma +m-n)}
\right. \nonumber\\ & ~~~~~~
-\frac{\cos (\pi  (\alpha -\beta +\gamma ))}{(-\alpha +\beta +l-m) (-\alpha -\gamma +l+n)}
+\frac{\cos (\pi  (\alpha -\beta +\gamma ))}{(-\alpha -\gamma +l+n) (-\beta +\gamma +m-n)}
\nonumber\\ & \left. ~~~~~~
-\frac{\cos (\pi  (\alpha +\beta +\gamma ))}{(-\alpha -\beta +l+m) (-\alpha -\gamma +l+n)}
-\frac{\cos (\pi  (\alpha +\beta +\gamma ))}{(-\alpha -\gamma +l+n) (-\beta -\gamma +m+n)}
\right)
\end{align}
The six series can be evaluated and the result reads 
\begin{align}
& \mathcal{J}(\alpha,\beta,\gamma)=
2\pi\, 
\nonumber\\&
\left(
\frac{\sqrt{\pi }\, \Gamma \left(\alpha +\frac{1}{2}\right) \Gamma \left(\gamma +\frac{1}{2}\right) \Gamma (-\alpha -\beta ) \cos (\pi  (\alpha +\beta -\gamma )) \Gamma (\alpha +\beta +\gamma +1)}{\sin (\pi  (\beta -\gamma )) \Gamma \left(\frac{1}{2}-\beta \right) \Gamma (\alpha +\gamma +1) \Gamma (\beta +\gamma +1)} +
\right. \nonumber\\&
-\frac{\sqrt{\pi }\, \Gamma \left(\alpha +\frac{1}{2}\right) \Gamma \left(\beta +\frac{1}{2}\right) \Gamma (-\alpha -\gamma ) \cos (\pi  (\alpha -\beta +\gamma )) \Gamma (\alpha +\beta +\gamma +1)}{\sin (\pi  (\beta -\gamma )) \Gamma \left(\frac{1}{2}-\gamma \right) \Gamma (\alpha +\beta +1) \Gamma (\beta +\gamma +1)} +
 \nonumber\\&
\frac{\sqrt{\pi }\, \Gamma \left(\beta +\frac{1}{2}\right) \Gamma \left(\gamma +\frac{1}{2}\right) \Gamma (-\alpha -\gamma ) \cos (\pi  (\alpha -\beta +\gamma )) \Gamma (\alpha +\beta +\gamma +1)}{\sin (\pi  (\alpha -\beta )) \Gamma \left(\frac{1}{2}-\alpha \right) \Gamma (\alpha +\beta +1) \Gamma (\beta +\gamma +1)} +
\nonumber\\&
-\frac{\sqrt{\pi }\, \Gamma \left(\alpha +\frac{1}{2}\right) \Gamma \left(\gamma +\frac{1}{2}\right) \Gamma (-\beta -\gamma ) \cos (\pi  (\alpha -\beta -\gamma )) \Gamma (\alpha +\beta +\gamma +1)}{\sin (\pi  (\alpha -\beta )) \Gamma \left(\frac{1}{2}-\beta \right) \Gamma (\alpha +\beta +1) \Gamma (\alpha +\gamma +1)} +
\nonumber\\&
-\frac{\Gamma \left(\alpha +\frac{1}{2}\right) \Gamma \left(\beta +\frac{1}{2}\right) \Gamma (-\alpha -\gamma ) \Gamma (-\beta -\gamma ) \cos (\pi  (\alpha +\beta +\gamma )) \Gamma (\alpha +\beta +\gamma +1)}{\sqrt{\pi } \Gamma \left(\frac{1}{2}-\gamma \right) \Gamma (\alpha +\beta +1)} +
\nonumber\\& \left.
-\frac{\Gamma \left(\beta +\frac{1}{2}\right) \Gamma \left(\gamma +\frac{1}{2}\right) \Gamma (-\alpha -\beta ) \Gamma (-\alpha -\gamma ) \cos (\pi  (\alpha +\beta +\gamma )) \Gamma (\alpha +\beta +\gamma +1)}{\sqrt{\pi } \Gamma \left(\frac{1}{2}-\alpha \right) \Gamma (\beta +\gamma +1)}
\right)
\end{align}
After non--trivial cancellations we obtain  
\begin{equation}
\label{J2}
\mathcal{J}(\alpha,\beta,\gamma)=
8\pi^{3/2} \,
\frac{\Gamma(\tfrac{1}{2}+\alpha) \Gamma(\tfrac{1}{2}+\beta) \Gamma(\tfrac{1}{2}+\gamma) \Gamma(1+\alpha+\beta+\gamma)}
{\Gamma(1+\alpha+\gamma)\Gamma(1+\beta+\gamma)\Gamma(1+\alpha+\beta)}
\end{equation}
We note that in all the six $\mathcal{J}$ integrals appearing in the combination (\ref{U2}), the parameters satisfy $\alpha+\beta+\gamma=-\tfrac{1}{2}+2\epsilon$. This allows to express for instance $\g$ as a function of $\a$ and $\b$, leading to expressions which depend only on the choice of two parameters. The six different contributions are then obtained by choosing $\alpha$ to be $u$, $u+\tfrac{1}{2}$ or $u+1$ and $\beta$ to be $v$, $v+\tfrac{1}{2}$ or $v+1$ in different combinations. What is relevant to observe is that in all the cases no ambiguous products of Gamma functions appear. 

Plugging expressions (\ref{J2}) with these choices of the parameters into the contribution (\ref{U2}), the result can be expressed as
\begin{align}
& U_{\rm f} = -\frac{2^{4\epsilon+2} \pi^{3-\epsilon} (1-2\epsilon) \Gamma \left( \frac{1}{2} + 2 \epsilon \right)}
{3\,\Gamma^3 \left( \frac12 - \epsilon \right) \Gamma\left( \frac12 + \epsilon \right)}
\Big[  G(1,\tfrac{1}{2})+G(\tfrac{1}{2},1)
\\& \phantom{dejo espacio yo tambien} \qquad \qquad 
+G(1,0) +G(0,1)+G(\tfrac{1}{2},0)+G(0,\tfrac{1}{2})\Big] \non 
\end{align}
in terms of the Mellin--Barnes integrals
\begin{align}
G(i,j)= &
\int \frac{du\,dv}{(2\pi i)^2}\, \Gamma (-u) \Gamma (-v) \Gamma \left(-\!u\!+\!\epsilon\! -\!\tfrac{1}{2}\right) \Gamma \left(-\!v\!+\!\epsilon\! -\!\tfrac{1}{2}\right)
\Gamma (u\!+\!v\!-\!2 \epsilon\! +\!2) \nonumber\\
&\times
\frac{\Gamma \left(u\!+\!v\!-\!\epsilon\! +\!\tfrac{3}{2}\right) \Gamma \left(i+u+\frac{1}{2}\right) \Gamma \left(j+v+\frac{1}{2}\right) \Gamma (-i-j-u-v+2 \epsilon )}{\Gamma \left(-i-u+2 \epsilon +\frac{1}{2}\right) \Gamma \left(-j-v+2 \epsilon +\frac{1}{2}\right) \Gamma (i+j+u+v+1)}
\end{align}
However, using the change of variables $(u\rightarrow -u-v+2\epsilon-2,\, v\rightarrow v)$, $(v \rightarrow -u-v+2\epsilon-2,\, u\rightarrow u)$ and $u\leftrightarrow v$, it is not difficult 
to show that the six $G(a,b)$ functions are indeed all equal. Therefore, we can write
\begin{align}
& U_{\rm f} = -\frac{2^{4\epsilon+3}\, \pi^{3-\epsilon} (1-2\epsilon) \Gamma \left(\frac{1}{2}+ 2 \epsilon \right)}
{\,\Gamma^3 \left( \frac12 - \epsilon \right) \Gamma\left( \frac12 + \epsilon \right)}
\ G(1,\tfrac{1}{2})
\end{align}
with
\begin{align}
G(1,\tfrac{1}{2})= &
\int \frac{du\,dv}{(2\pi i)^2}\, \Gamma (-u) \Gamma (-v) \Gamma \left(-\!u\!+\!\epsilon\! -\!\tfrac{1}{2}\right) \Gamma \left(-\!v\!+\!\epsilon\! -\!\tfrac{1}{2}\right)
\Gamma (u\!+\!v\!-\!2 \epsilon\! +\!2) \nonumber\\
&\times
\frac{\Gamma \left(u\!+\!v\!-\!\epsilon\! +\!\tfrac{3}{2}\right) \Gamma \left(u+\frac{3}{2}\right) \Gamma (v+1) \Gamma \left(-u-v+2 \epsilon -\frac{3}{2}\right)}{\Gamma \left(u+v+\frac{5}{2}\right) \Gamma \left(-u+2 \epsilon -\frac{1}{2}\right) \Gamma (2 \epsilon -v)}
\end{align}
First we expand the integral in powers of $\epsilon$ up to finite order, which gives a one--fold and a two--fold Mellin--Barnes integrals. The former integral reads
\begin{align}
\frac{\sqrt{\pi }}{\epsilon } \int \frac{du}{2\pi i}\, \Gamma \left(-u-\frac{1}{2}\right) \Gamma (-u) \Gamma (u+1) \Gamma \left(u+\frac{3}{2}\right) \left(\epsilon  \psi ^{(0)}\left(-u-\frac{1}{2}\right)-\epsilon  \psi ^{(0)}(u+1)
\right. \nonumber\\ \left.
-2 \epsilon  \psi ^{(0)}\left(u+\frac{3}{2}\right)+\gamma_E  \epsilon +1\right)
\end{align}
where $\psi ^{(0)}$ is the digamma function defined in (\ref{gammader}). It can be evaluated applying Barnes first lemma and we obtain
\begin{equation}
\frac{\pi^{3/2}}{2}\, \left( \frac{1}{\epsilon} - 2 + 3 \gamma_E + 4 \log 2 \right)
\end{equation}
The latter integral is
\begin{equation}
\int \frac{du\,dv}{(2\pi i)^2}\, \frac{\Gamma^* \left(-u-\frac{3}{2}\right) \Gamma \left(u+\frac{3}{2}\right) \Gamma (u+2) \Gamma \left(-v-\frac{1}{2}\right) \Gamma (v+1) \Gamma \left(u-v+\frac{3}{2}\right) \Gamma (v-u)}{\Gamma \left(u+\frac{5}{2}\right)}
\end{equation} 
where, according to the notations of \cite{Smirnov}, asterisks denote how many of the first right (left) poles of the Gamma functions have to be considered left (right).

With a change of variables and applying Barnes first lemma this can be reduced to a one--fold integral
\begin{equation}
\frac{\pi}{2} \int \frac{du}{2\pi i}\, \Gamma^* \left(-u-\frac{3}{2}\right) \Gamma \left(-u-\frac{1}{2}\right) \Gamma \left(u+\frac{3}{2}\right) \Gamma (u+2)
\end{equation}
The remaining integral is evaluated by lemma $(D.3)$ of \cite{Smirnov} and gives
\begin{equation}
\pi^{3/2} \left( -1+\log 2 \right)
\end{equation}
Summing all the contributions we obtain
\begin{equation}
G(1,\tfrac{1}{2})= \frac{\pi^{3/2}}{2}\, \left( \frac{1}{\epsilon} - 4 + 3 \gamma_E + 6 \log 2 \right)
\end{equation}
and, consequently
\begin{equation}
U_f = 4\pi^3
\left(-\frac{1}{\epsilon}+6+\gamma_E-2\log 2 +\log \pi \right)
\end{equation}

\section{Evaluation of the contracted integrals}
\label{contracted}

Here we solve the parametric integrals appearing in eq. (\ref{eq:intfcontr}) for the contracted integrals $C_{\rm f}$ from diagram \ref{2loop}(f).  
 
\vskip 15pt
\noindent 
1) We start with the first integral
\begin{align}\label{eq:Icontr1}
& 
I_{(\ref{eq:Icontr1})} =
-2^{4 \epsilon -3} \int _0^{2 \pi } \!\!\!\!d\tau_1  \int _0^{\tau_1} \!\!\!\! d\tau_2 \int _0^{\tau_2}\!\!\!\!d\tau_3\,  \frac{\sin (\t_{13})}{ \left(\sin \left(\frac{\t_{12}}{2}\right) \sin \left(\frac{\t_{23}}{2}\right)\right)^{2-2 \epsilon}}
\end{align}
where the overall factor in front has been introduced for later convenience. 

Applying the procedure of the previous Section, in particular already removing the imaginary part, we turn it into a combination of series
\begin{align}
I_{(\ref{eq:Icontr1})}
= & 
2 \pi  \sin (2 \pi  \epsilon ) \Big( S_{1,1,0}[2-\epsilon, 2-\epsilon] - S_{1,1,0}[-\epsilon, -\epsilon ] \Big) +
\nonumber\\&
+
2 \left( 1-\cos (2 \pi  \epsilon ) \right) 
\Big(
S_{1,2,0}[2-\epsilon, 2-\epsilon] - S_{1,2,0}[-\epsilon, -\epsilon]
\Big)
\end{align}
where we have defined
\begin{equation}
\label{doubleseries}
S_{m,n,p}[a,b] = \sum_{u=0}^{\infty} \sum_{v=0}^{\infty}\, 
\frac{\Gamma (u-2 \epsilon +2) \Gamma (v-2 \epsilon +2)}{\Gamma (u+1) \Gamma (v+1) \Gamma^2 (2-2 \epsilon )} \frac{1}{(u+a)^m (v+b)^n (u+v+a+b)^p}
\end{equation}
Using
\begin{align}
S_{1,1,0}[a,b] &= 
\frac{\Gamma (a) \Gamma (b) \Gamma^2 (-1+2 \epsilon)}{\Gamma (a+2 \epsilon -1) \Gamma (b+2 \epsilon -1)}
\nonumber\\
S_{1,2,0}[a,b] &= S_{2,1,0}[b,a] = \frac{\Gamma (a) \Gamma (b) \Gamma^2 (-1+2 \epsilon) \left( \psi ^{(0)}(b+2 \epsilon -1) - \psi ^{(0)}(b) \right)}{\Gamma (a+2 \epsilon -1) \Gamma (b+2 \epsilon -1)}
\end{align}
after some algebra we obtain
\begin{equation}
I_{(\ref{eq:Icontr1})}
=
\frac{8 \pi ^2 (\epsilon -1) \epsilon  (2 \epsilon -1) \Gamma^2 (-1+2 \epsilon)}{\Gamma^4 (1+ \epsilon)}
\end{equation}

\vskip 20pt
\noindent 
2) We then move to the second integral appearing in eq. (\ref{eq:intfcontr})
\begin{equation}
\label{eq:Icontr2}
I_{(\ref{eq:Icontr2})} = 
2^{4 \epsilon -3} \,
\int _0^{2 \pi } \!\!\!\!d\tau_1 \int _0^{\tau_1} \!\!\!\!d\tau_2 \int _0^{\tau_2} \!\!\!\!d\tau_3\,  
\frac{\sin (\t_{12})}{\left(\sin \left(\frac{\t_{13}}{2}\right) \sin \left(\frac{\t_{23}}{2}\right)\right)^{2-2 \epsilon}}
\end{equation}
Turning it into series and discarding the imaginary part gives
\begin{align}
\label{second}
I_{(\ref{eq:Icontr2})}
= &
2 \pi  \sin (2 \pi  \epsilon ) \Big( S_{1,0,1}[-\epsilon, 2-\epsilon ] - S_{1,0,1}[2-\epsilon, -\epsilon] \Big) +
\nonumber\\&
+
\left( 1-\cos (2 \pi  \epsilon ) \right) 
\Big(
S_{1,2,0}[2-\epsilon, -\epsilon] - S_{2,1,0}[2-\epsilon, -\epsilon]
\Big)
\end{align}
Summing the $S_{1,0,1}$ series we find
\begin{equation}
S_{1,0,1}
= S_{1,1,0}[a,b]
-\frac{\Gamma (b) \Gamma (-1+2 \epsilon) \Gamma (a+b)}{\G(b+1) \G(a+b+2\e-1)} \, 
{}_3F_2\left[
             \begin{array}{c}
             b ,a+b ,2-2 \epsilon \\
             b+1,a+b+2 \epsilon -1 \\
             \end{array};1
             \right]
\end{equation}
Therefore, the final result for this integral reads
\begin{align}
I_{(\ref{eq:Icontr2})} 
= &
\,\, \frac{2 \pi ^2}{2-\epsilon } \; 
{}_3F_2\left[
             \begin{array}{c}
             2-2 \epsilon ,2-2 \epsilon ,2-\epsilon \\
             1,3-\epsilon \\
             \end{array};1
             \right]
+ \nonumber\\&
+ \frac{2 \pi ^2}{\epsilon} \; 
{}_3F_2\left[
             \begin{array}{c}
             2-2 \epsilon ,2-2 \epsilon ,-\epsilon \\
             1,1-\epsilon \\
             \end{array};1
             \right]
+\frac{4 \pi  \sin (\pi  \epsilon ) \Gamma (2-\epsilon ) \Gamma (2 \epsilon ) \Gamma (-1+ 2 \epsilon)}{\Gamma^3 (1+\epsilon)}
\end{align}

\vskip 15pt
It turns out that expanding the third integrand in eq. (\ref{eq:intfcontr})   
\begin{equation}
I_{(\ref{eq:Icontr2})b}
=
2^{4 \epsilon -3} \int _0^{2 \pi }\!\!\!\! d\tau_1\int _0^{\tau_1}\!\!\!\! d\tau_2 \int _0^{\tau_2} \!\!\!\!d\tau_3\, \frac{\sin (\t_{23})}{\left(\sin \left(\frac{\t_{12}}{2}\right) \sin \left(\frac{\t_{13}}{2}\right)\right)^{2-2 \epsilon}}
\end{equation}
and performing the integrations term by term we obtain exactly the result (\ref{second}). Therefore, the third integral simply contributes in doubling the previous result.

\vskip 20pt
\noindent 
3) Next we turn to the sum of the integrals
\begin{align}
\label{eq:Icontr3}
& I_{(\ref{eq:Icontr3})} =  
-2^{4 \epsilon -1}  \int _0^{2 \pi } \!\!\!\!d\tau_1 \int _0^{\tau_1} \!\!\!\!d\tau_2 \int _0^{\tau_2} \!\!\!\!d\tau_3 \; \times  
\\
& \left( \frac{\sin \left(\frac{\t_{23}}{2}\right)}{\left(\sin \left(\frac{\t_{12}}{2}\right)\sin \left(\frac{\t_{13}}{2}\right)\right)^{1-2 \epsilon}} 
+ \frac{\sin \left(\frac{\t_{13}}{2}\right)}{\left(\sin \left(\frac{\t_{12}}{2}\right)\sin \left(\frac{\t_{23}}{2}\right)\right)^{1-2 \epsilon}}
+ \frac{\sin \left(\frac{\t_{12}}{2}\right)}{\left(\sin \left(\frac{\t_{13}}{2}\right) \sin \left(\frac{\t_{23}}{2}\right)\right)^{1-2\epsilon}} \right)
\non \\
\non
\end{align}
corresponding to the second line in eq. (\ref{eq:intfcontr}). 

Although these integrals can be solved separately with the technique previously described, it turns out to be more convenient to consider their sum, as it leads to a considerable technical simplification.   

First of all, given the particular domain of integration, the arguments $\tau_{ij}/2$ of the trigonometric functions are always bounded between 0 and $\pi$. This allows to trade any  expression  $\sin \left(\tfrac{\t_{ij}}{2}\right) $ for $\sin \left(\tfrac{|\t_{ij}|}{2}\right) $. Now, the crucial observation is that, when rewritten in this form, the sum in (\ref{eq:Icontr3}) turns out to be totally symmetric under any exchange of  the $\tau_i$ parameters. Therefore, we can symmetrize the integration contours as
\begin{equation}
\int _0^{2 \pi } d\tau_1 \int _0^{\tau_1} d\tau_2 \int _0^{\tau_2} d\tau_3
\rightarrow \frac{1}{3!}
\int _0^{2 \pi } d\tau_1 \int _0^{2\pi} d\tau_2 \int _0^{2\pi} d\tau_3  
\end{equation}
As a consequence, the three separate integrals in eq. (\ref{eq:Icontr3}) become the same integral. Taking into account an overall factor $3$, we can then write  
\begin{equation}
I_{(\ref{eq:Icontr3})} = -2^{4 \epsilon -2} \int _0^{2 \pi } \!\!\!\!d\tau_1 \int _0^{2\pi} \!\!\!\!d\tau_2 \int _0^{2\pi} \!\!\!\!d\tau_3 \, \frac{\sin \left(\frac{|\t_{23}|}{2}\right)}{\left(\sin \left(\frac{|\t_{12}|}{2}\right)\sin \left(\frac{|\t_{13}|}{2}\right)\right)^{1-2 \epsilon}}
\end{equation}
Performing the shift of integration variables $\tau_3\rightarrow\tau_3+\tau_1$ and  $\tau_2\rightarrow\tau_2+\tau_1$ and exploiting the $2\pi$-periodicity of the integrand as a function of $\tau_2, \tau_3$ we finally arrive to  
\begin{equation}
I_{(\ref{eq:Icontr3})} = -2^{4 \epsilon -1} \pi\, \int _0^{2 \pi }\!\!\!\!d\tau_2\int _{0}^{2\pi}\!\!\!\!d\tau_3\,
\frac{\sin \left(\frac{|\t_{23}|}{2}\right)}{\left(\sin \left(\frac{|\t_{2}|}{2}\right)\sin \left(\frac{|\t_{3}|}{2}\right)\right)^{1-2 \epsilon}}
\end{equation}
The remaining integrals can be solved with the method of Appendix \ref{sec:method} and the result is simply
\begin{equation}
I_{(\ref{eq:Icontr3})}
=
-\frac{\pi ^{3/2} 2^{4 \epsilon +1} \Gamma \left(\frac{1}{2} + 2 \epsilon \right)}{\epsilon ^2 \Gamma (2 \epsilon )}
\end{equation}

\vskip 20pt
\noindent 
4) Finally we consider the following combinations
\begin{align}
\label{eq:Icontr4}
& I_{(\ref{eq:Icontr4})} = 
2^{4 \epsilon -2} \,  \int _0^{2 \pi } \!\!\!\!d\tau_1 \int _0^{\tau_1} \!\!\!\!d\tau_2 \int _0^{\tau_2}\!\!\!\! d\tau_3 \; \times 
\\&
\left( \frac{\cos \left(\frac{\t_{23}}{2}\right)}{ \sin ^{2-2 \epsilon }\left(\frac{\t_{12}}{2}\right) \sin ^{1-2 \epsilon}\left(\frac{\t_{13}}{2}\right)} +
\frac{\cos \left(\frac{\t_{23}}{2}\right)}{ \sin ^{1-2 \epsilon}\left(\frac{\t_{12}}{2}\right) \sin ^{2-2 \epsilon}\left(\frac{\t_{13}}{2}\right)}
- \frac{\cos \left(\frac{\t_{13}}{2}\right)}{ \sin ^{1-2 \epsilon}\left(\frac{\t_{12}}{2}\right) \sin ^{2-2 \epsilon}\left(\frac{\t_{23}}{2}\right)}\right)
\non \\
\non
\end{align}
and 
\begin{align}
\label{eq:Icontr5}
& I_{(\ref{eq:Icontr5})} =
2^{4 \epsilon -2} \,  \int _0^{2 \pi } \!\!\!\!d\tau_1 \int _0^{\tau_1} \!\!\!\! d\tau_2 \int _0^{\tau_2}\!\!\!\!d\tau_3 \; \times
\\& 
 \left( \frac{\cos \left(\frac{\t_{12}}{2}\right)}{ \sin ^{2-2 \epsilon}\left(\frac{\t_{13}}{2}\right) \sin ^{1-2 \epsilon}\left(\frac{\t_{23}}{2}\right)}
+ \frac{\cos \left(\frac{\t_{12}}{2}\right)}{ \sin ^{1-2 \epsilon}\left(\frac{\t_{13}}{2}\right) \sin ^{2-2 \epsilon}\left(\frac{\t_{23}}{2}\right)}
-\frac{\cos \left(\frac{\t_{13}}{2}\right)}{ \sin ^{2-2 \epsilon}\left(\frac{\t_{12}}{2}\right) \sin ^{1-2 \epsilon}\left(\frac{\t_{23}}{2}\right)}   \right)
\non
\end{align}
appearing in the third and fourth lines of eq. (\ref{eq:intfcontr}), respectively. 
Expanding the integrands in power series, it is easy to realize that these integrals give pairwise the same results. Therefore, it is sufficient to evaluate one of the two combinations. 

In principle, each integral in (\ref{eq:Icontr4}) could be performed separately. However, once again the sum turns out to be far simpler to compute than the individual pieces, this time due to 
considerable cancellations which involve the most difficult parts of the series.

After a quite cumbersome algebra and many intermediate cancellations, the final expression in terms of series reads
\bea
I_{(\ref{eq:Icontr4})} &=&
2 \pi  \sin (2 \pi  \epsilon ) \Big( \tilde{S}_{1,1}[1-\epsilon, 1-2\epsilon ] + \tilde{S}_{1,1}[-\epsilon, 3-2\epsilon]  
\nonumber\\   
&& \qquad \qquad \qquad \qquad -\tilde{S}_{1,1}[1-\epsilon, 3-2\epsilon ] - \tilde{S}_{1,1}[-\epsilon, 1-2\epsilon ] \Big) 
\nonumber\\
&+& \left( 1-\cos (2 \pi  \epsilon ) \right) 
\Big( -2\, \tilde{S}_{1,2}[-\epsilon, 1-2\epsilon] -2\, \tilde{S}_{1,2}[1-\epsilon, 3-2\epsilon]
 \nonumber\\  
&& \qquad \qquad \qquad \qquad -\tilde{S}_{2,1}[1-\epsilon, 3-2\epsilon] - \tilde{S}_{2,1}[-\epsilon, 1-2\epsilon]
\Big) 
\nonumber\\
&+& \left( 1+\cos (2 \pi  \epsilon ) \right) 
\Big( - 2\, \tilde{S}_{1,2}[1-\epsilon, 1-2\epsilon] - 2\, \tilde{S}_{1,2}[-\epsilon, 3-2\epsilon]  
\nonumber\\  
&& \qquad \qquad \qquad \qquad + \tilde{S}_{2,1}[-\epsilon, 3-2\epsilon] + \tilde{S}_{2,1}[1-\epsilon, 1-2\epsilon]
\Big)
\nonumber\\
&+& S_1 + S_2 + S_3 + S_4
\eea
where we have defined 
\bea
S_1 &=& \sum_{u=0}^{\infty} \sum_{v=0}^{\infty}\, \frac{16\, \Gamma (u-2 \epsilon +1) \Gamma (v-2 \epsilon +2)}{\Gamma (u+1) \Gamma (v+1) \Gamma (1-2 \epsilon ) \Gamma (2-2 \epsilon ) (2 u-2 v-1) (2 v-2 \epsilon +1)^2}
\nonumber\\
\non \\
&=& - \frac{16 \sqrt{\pi } \Gamma (2 \epsilon ) \, 
{}_4F_3\left[
             \begin{array}{c}
             \frac{3}{2}-2 \epsilon ,2-2 \epsilon ,\frac{1}{2}-\epsilon ,\frac{1}{2}-\epsilon \\
             \frac{3}{2},\frac{3}{2}-\epsilon ,\frac{3}{2}-\epsilon \\
             \end{array};1
             \right]}{(2 \epsilon -1)^2 \Gamma \left( -\frac{1}{2} + 2\epsilon  \right)} \non \\
\eea
\bea
S_2 &=& \sum_{u=0}^{\infty} \sum_{v=0}^{\infty}\, \frac{16\, \Gamma (u-2 \epsilon +1) \Gamma (v-2 \epsilon +2)}{\Gamma (u+1) \Gamma (v+1) \Gamma (1-2 \epsilon ) \Gamma (2-2 \epsilon ) (2 u-2 v-1) (2 v-2 \epsilon +3)^2}
\nonumber\\
\non \\
&=& -\frac{16 \sqrt{\pi } \Gamma (2 \epsilon ) \, 
{}_4F_3\left[
             \begin{array}{c}
             \frac{3}{2}-2 \epsilon ,2-2 \epsilon ,\frac{3}{2}-\epsilon ,\frac{3}{2}-\epsilon \\
             \frac{3}{2},\frac{5}{2}-\epsilon ,\frac{5}{2}-\epsilon \\
             \end{array};1
             \right]}{(2 \epsilon -3)^2 \Gamma \left(-\frac{1}{2} + 2 \epsilon \right)} \non \\
\eea
\bea
S_3 &=& - \sum_{u=0}^{\infty} \sum_{v=0}^{\infty}\, \frac{4\, \Gamma (u-2 \epsilon +1) \Gamma (v-2 \epsilon +2) \cos (2 \pi  \epsilon )}{\Gamma (u+1) \Gamma (v+1) \Gamma (1-2 \epsilon ) \Gamma (2-2 \epsilon ) (u-\epsilon )^2 (2 u+2 v-4 \epsilon +3)}
\nonumber\\
\non \\
&=& \frac{4 \cos (2 \pi  \epsilon ) \Gamma \left(\tfrac52-2 \epsilon\right) \Gamma (-1 + 2 \epsilon) \, 
{}_4F_3\left[
             \begin{array}{c}
             1-2 \epsilon ,\frac{3}{2}-2 \epsilon ,-\epsilon ,-\epsilon \\
             \frac{1}{2},1-\epsilon ,1-\epsilon \\
             \end{array};1
             \right]}{\sqrt{\pi } \epsilon ^2 (4 \epsilon -3)} \non \\
\eea
\bea
S_4 &=& - \sum_{u=0}^{\infty} \sum_{v=0}^{\infty}\, \frac{4\, \Gamma (u-2 \epsilon +1) \Gamma (v-2 \epsilon +2) \cos (2 \pi  \epsilon )}{\Gamma (u+1) \Gamma (v+1) \Gamma (1-2 \epsilon ) \Gamma (2-2 \epsilon ) (u-\epsilon +1)^2 (2 u+2 v-4 \epsilon +3)}
\nonumber\\
\non \\
&=& \frac{4 \cos (2 \pi  \epsilon ) \Gamma \left(\tfrac52-2 \epsilon\right) \Gamma (-1 +2 \epsilon) \, 
{}_4F_3\left[
             \begin{array}{c}
             1-2 \epsilon ,\frac{3}{2}-2 \epsilon ,1-\epsilon ,1-\epsilon \\
             \frac{1}{2},2-\epsilon ,2-\epsilon \\
             \end{array};1
             \right]}{\sqrt{\pi } (\epsilon -1)^2 (4 \epsilon -3)} \non \\
\eea
and 
\beq
\tilde{S}_{m,n}[a,b] = \sum_{u=0}^{\infty} \sum_{v=0}^{\infty}\, \frac{2 \Gamma (u-2 \epsilon +1) \Gamma (v-2 \epsilon +2)}{\Gamma (u+1) \Gamma (v+1)\Gamma (1-2 \epsilon ) \Gamma (2-2 \epsilon )}\, \frac{1}{(u+a)^m (2v+b)^n}
\eeq
\\
\noindent
In particular, we need the following sums
\begin{align}
\tilde{S}_{1,1}[a,b] &= \frac{2 \Gamma (a) \Gamma \left(\frac{b+2}{2}\right) \Gamma (2 \epsilon ) \Gamma (-1 +2 \epsilon)}{b \Gamma (a+2 \epsilon ) \Gamma \left(\frac{b}{2}+2 \epsilon -1\right)}
\nonumber\\
\tilde{S}_{1,2}[a,b] &= \frac{2 (\epsilon -1)\Gamma (-2 +2\epsilon) \Gamma (2 \epsilon )  \Gamma (a) \Gamma \left(\frac{b+2}{2}\right) \left(\psi ^{(0)}\left(\frac{b}{2}+2 \epsilon -1\right)-\psi ^{(0)}\left(\frac{b}{2}\right)\right)}{b \Gamma (a+2 \epsilon ) \Gamma \left(\frac{b}{2}+2 \epsilon -1\right)}
\nonumber\\
\tilde{S}_{2,1}[a,b] &= \frac{2 \Gamma (2 \epsilon ) \Gamma (-1+2 \epsilon) \Gamma (a) \Gamma \left(\frac{b+2}{2}\right) \left(\psi ^{(0)}(a+2 \epsilon )-\psi ^{(0)}(a)\right)}{b \Gamma (a+2 \epsilon ) \Gamma \left(\frac{b}{2}+2 \epsilon -1\right)}
\end{align}
\\
\noindent
from which we obtain
\begin{align}
I_{(\ref{eq:Icontr4})} & =   
\frac{4 \cos (2 \pi  \epsilon ) \Gamma \left(\tfrac52-2 \epsilon\right) \Gamma (-1 + 2 \epsilon)}{\sqrt{\pi }(4\epsilon -3)\epsilon ^2}\, 
{}_4F_3\left[
             \begin{array}{c}
             1-2 \epsilon ,\frac{3}{2}-2 \epsilon ,-\epsilon ,-\epsilon \\
             \frac{1}{2},1-\epsilon ,1-\epsilon \\
             \end{array};1
             \right]
\nonumber\\&
+ \frac{4 \cos (2 \pi  \epsilon ) \Gamma \left(\tfrac52-2 \epsilon\right) \Gamma (-1 + 2 \epsilon)}{\sqrt{\pi }(4\epsilon -3)(\epsilon -1)^2}\, 
{}_4F_3\left[
             \begin{array}{c}
             1-2 \epsilon ,\frac{3}{2}-2 \epsilon ,1-\epsilon ,1-\epsilon \\
             \frac{1}{2},2-\epsilon ,2-\epsilon \\
             \end{array};1
             \right]
\nonumber\\&
-\frac{16 \sqrt{\pi } \Gamma (2 \epsilon ) \, 
{}_4F_3\left[
             \begin{array}{c}
             \frac{3}{2}-2 \epsilon ,2-2 \epsilon ,\frac{1}{2}-\epsilon ,\frac{1}{2}-\epsilon \\
             \frac{3}{2},\frac{3}{2}-\epsilon ,\frac{3}{2}-\epsilon \\
             \end{array};1
             \right]}{(2 \epsilon -1)^2 \Gamma \left(-\frac{1}{2} + 2 \epsilon \right)}
\nonumber\\&
-\frac{16 \sqrt{\pi } \Gamma (2 \epsilon ) \, 
{}_4F_3\left[
             \begin{array}{c}
             \frac{3}{2}-2 \epsilon ,2-2 \epsilon ,\frac{3}{2}-\epsilon ,\frac{3}{2}-\epsilon \\
             \frac{3}{2},\frac{5}{2}-\epsilon ,\frac{5}{2}-\epsilon \\
             \end{array};1
             \right]}{(2 \epsilon -3)^2 \Gamma \left(-\frac{1}{2}+ 2 \epsilon  \right)}
\nonumber\\&
+\frac{\pi ^2 16^{\epsilon } \csc^2(2 \pi  \epsilon ) \left(\sin^2(2 \pi  \epsilon )-2 \cos (2 \pi  \epsilon )\right)}{\epsilon }
\end{align}
Finally, combining the results of this Section we obtain expression (\ref{intfcontr}) for the contracted integrals. 
The hypergeometric functions do not have a well--defined expansion around $\e =0$ and analytic continuation has to be performed prior expanding in powers of $\epsilon$. We explained the proper analytical continuation in Section (\ref{sec:expansions}).

\section{Useful series}\label{app:series}

In order to determine explicitly finite order terms in the $\e$--expansions of extended hypergeometric functions ${}_4 F_3$ appearing in Section \ref{sec:expansions} we need know the sum of several series. Most of them can be found in the literature, but there are few that cannot. For the last ones we have performed an explicit evaluation. The results are listed below.

We use the standard definition for the derivatives of the $\Gamma$ function
\begin{equation}
\label{gammader}
\psi^{(n)}(z)=\frac{d^{n+1}}{dz^{n+1}}\log\Gamma(z)
\end{equation}

The following results are useful
\begin{equation}
\sum\limits_{n=1}^{\infty}\frac{\psi^{(1)}(\tfrac{1}{2}+n)}{n(n-\tfrac{1}{2})}=
4\pi ^2 \log 2-21 \zeta (3)
\end{equation}
\begin{equation}
\sum\limits_{n=1}^{\infty}\frac{\psi^{(1)}(-\tfrac{1}{2}+n)}{n(n+\tfrac{1}{2})}=
7 \zeta (3)+4-16 \log 2+\pi ^2 (3-4 \log 2)
\end{equation}
\begin{equation}
\sum\limits_{n=2}^{\infty}\frac{\psi^{(1)}(-1+n)}{n(n-\tfrac{1}{2})}=
3 \zeta (3)-14+16 \log 2
\end{equation}
\begin{equation}
\sum\limits_{n=1}^{\infty}\frac{\psi^{(1)}(1+n)}{n(n+\tfrac{1}{2})}=
\frac{2}{3}\pi ^2-5 \zeta (3)
\end{equation}

\end{document}